\documentclass[pdflatex,sn-nature]{sn-jnl}

\usepackage{graphicx}%
\usepackage{multirow}%
\usepackage{amsmath,amssymb,amsfonts}%
\usepackage{amsthm}%
\usepackage{mathrsfs}%
\usepackage[title]{appendix}%
\usepackage{xcolor}%
\usepackage{textcomp}%
\usepackage{manyfoot}%
\usepackage{booktabs}%
\usepackage{algorithm}%
\usepackage{algorithmicx}%
\usepackage{algpseudocode}%
\usepackage{listings}%
\usepackage{xcolor}
\usepackage{float}
\usepackage{booktabs}
\usepackage{placeins}
\usepackage{siunitx}

\usepackage{fancyhdr}
\usepackage{tocloft}
\usepackage{caption}

\fancypagestyle{supp}{%
  \fancyhf{}%
  \fancyhead[R]{\thepage}%
}

\definecolor{tocred}{RGB}{150,0,0}

\providecommand{\phantomsection}{}   
\newcommand{\suppheading}[1]{%
  \phantomsection
  \section*{#1}%
  \addcontentsline{toc}{section}{#1}%
}

\theoremstyle{thmstyleone}%
\theoremstyle{thmstyletwo}%

\theoremstyle{thmstylethree}%

\begin{document}

\title[Article Title]{Programmable Preparation of Topological Entangled States}

\author[1]{\fnm{M. Javad}\sur{Zakeri}}
\author[1]{\fnm{Alaric}\sur{Pan}}

\author[1]{\fnm{Darren D}\sur{Hudson}}

\author*[1]{\fnm{Andrea}\sur{Blanco-Redondo}}\email{Andrea.BlancoRedondo@ucf.edu}

\affil[1]{CREOL, The College of Optics and Photonics, University of Central Florida, Orlando, Florida 32816, USA}



\abstract{
Programmable preparation of entangled quantum states is a central requirement for scalable quantum information processing, quantum networking and photonic quantum computing. Topological photonic platforms offer an attractive route towards robust quantum-state generation by encoding information in symmetry-protected optical modes, yet the generated entangled states are typically dictated by the lattice geometry and therefore lack programmability. Here we demonstrate programmable preparation of topological entangled states by inverse engineering the nonlinear optical pump profile of an integrated photonic lattice. Our approach enables deterministic control over both the dimensionality and modal structure of the generated entangled state while preserving its topological character. We experimentally realize silicon photonic topological superlattices in which the nonlinear generation pathway itself becomes programmable via tailored excitations. The superlattices host a multiband manifold of co-localized topological interface modes, while coherently controlled inputs synthesize pump superpositions within this protected basis. 
We showcase this principle by selectively addressing topological biphoton eigenstates, steering correlations to produce a desired target state in a high-dimensional Hilbert space, and inverse-designing pump states that generate targeted Schmidt dimensionalities. Site-resolved two-photon correlation measurements remain reproducible across nominally identical fabricated devices, demonstrating robustness to realistic nanofabrication-induced disorder. 
Programmable preparation of topological entangled states establishes a pathway towards exploiting topological photonic modes as application-specific quantum resources, enabling tailored input states for high-dimensional quantum communication, quantum networking and photonic quantum computing. 
}


\maketitle

\section*{Main}\label{sec1}

Encoding quantum information in high-dimensional Hilbert spaces unlocks protocols with enhanced information capacity, robustness, and functionality beyond conventional qubit-based implementations \cite{erhard2020advances}. Photons are particularly well-suited to this high-dimensional regime given that most of their degrees of freedom to encode quantum information - path, temporal, orbital angular momentum, frequency, transverse spatial modes, etc. - are largely multimode \cite{malik2016multi}. Integrated photonics provides a scalable route to exploiting these degrees of freedom \cite{wang2020integrated}, with chip-based sources having demonstrated frequency- and path-entangled qudits of increasing dimensionality \cite{kues2017chip,wang2018multidimensional}. 


In the last decade, topology has emerged as a powerful route to establish a resilient modal basis for photonic entanglement \cite{blanco2019topological,JalaliMehrabad2023,Gao2024, Hashemi2025}. Topological photonic platforms can be engineered to support interface states that are spatially localized, spectrally isolated, and resistant to imperfections \cite{Lu2014TopologicalPhotonics, Ozawa2019TopologicalPhotonics,price2022roadmap}. Leveraging nonlinear optical effects - specifically spontaneous four-wave mixing (SFWM) - to generate correlated photon pairs in these topologically protected modes has been the workhorse of topological quantum photonics research, enabling spatially \cite{blanco-redondo2016cleo,blanco2018topological} and spectrally robust biphoton correlations \cite{mittal2018topological,WANG2022100003}, as well as resilient path \cite{wang2019topologically}, polarization \cite{WANG2022100003}, Einstein-Podolsky-Rosen \cite{dai2022topologically}, and chirality entanglement \cite{Santhosh2026chirality}. While most topological quantum sources operate on a reduced modal basis of one or two modes, a route for high-dimensional entanglement based on multiband topological superlattices has recently been established \cite{ZakeriCLEO2024, Zakeri2026}. Multiband topological superlattices \cite{PhysRevA.98.043838,Wang2021TopologicalSuperlattice} support a manifold of co-localized interface modes that form a high-dimensional topological basis for photonic entanglement when combined with appropriate lattice excitation and consequent leveraging of the optical nonlinearity in the structure. 


The remaining challenge to harness these resilient high-dimensional quantum states for quantum information processing is to gain precise control over the preparation of topologically protected entangled states. In a multimode system, the nonlinear optical process of SFWM can occur between any combination of four modes, however, its strength depends on the overlap of the excitation with such modes profiles and between the four mode profiles themselves \cite{Doyle2022biphoton}. Therefore, we propose that coherent pump shaping within the topological-mode manifold provides direct control over the signal--idler wavefunction and thus over the shape and dimensionality of the generated quantum states.\\ 
Here we demonstrate engineered preparation of topological entangled states using this principle. Specifically, we design and fabricate topological superlattices composed of silicon photonic waveguides with coherently controlled pump inputs that allow us to synthesize superpositions of localized interface modes and drive SFWM into selected signal--idler channels. To showcase the quantum state preparation capabilities of the system, we experimentally demonstrate three scenarios for the generation of quantum states: selective excitation of a specific eigenmode to show mode-specific generation of topological biphoton states; the generation of a family of coherent entangled states targeting the maximization of the weight of an arbitrarily chosen mode within the Hilbert space by pump engineering; and the targeting of a specific dimensionality (Schmidt number) that can be achieved with a variety of different excitation profiles. The quantum states at the output of the silicon photonic structures are characterized via measurements of the biphoton correlation maps, also referred to as joint spatial intensities (JSIs), and evaluated against phase-resolved propagation simulations. The characteristics of the quantum states are analyzed using projection onto the modal basis and Schmidt decomposition. Quantum measurements across nominally identical devices preserve their most important features, showcasing resilience to nanofabrication tolerances. These results establish a method for the controlled preparation of topological quantum states with arbitrary shapes and dimensionality, which opens the door for using topological entangled states as computational resources.

\section*{Results}
\label{sec2}

\textbf{A programmable topological quantum photonics platform}\\

\begin{figure}[H]
\centering
\includegraphics[width=\textwidth]{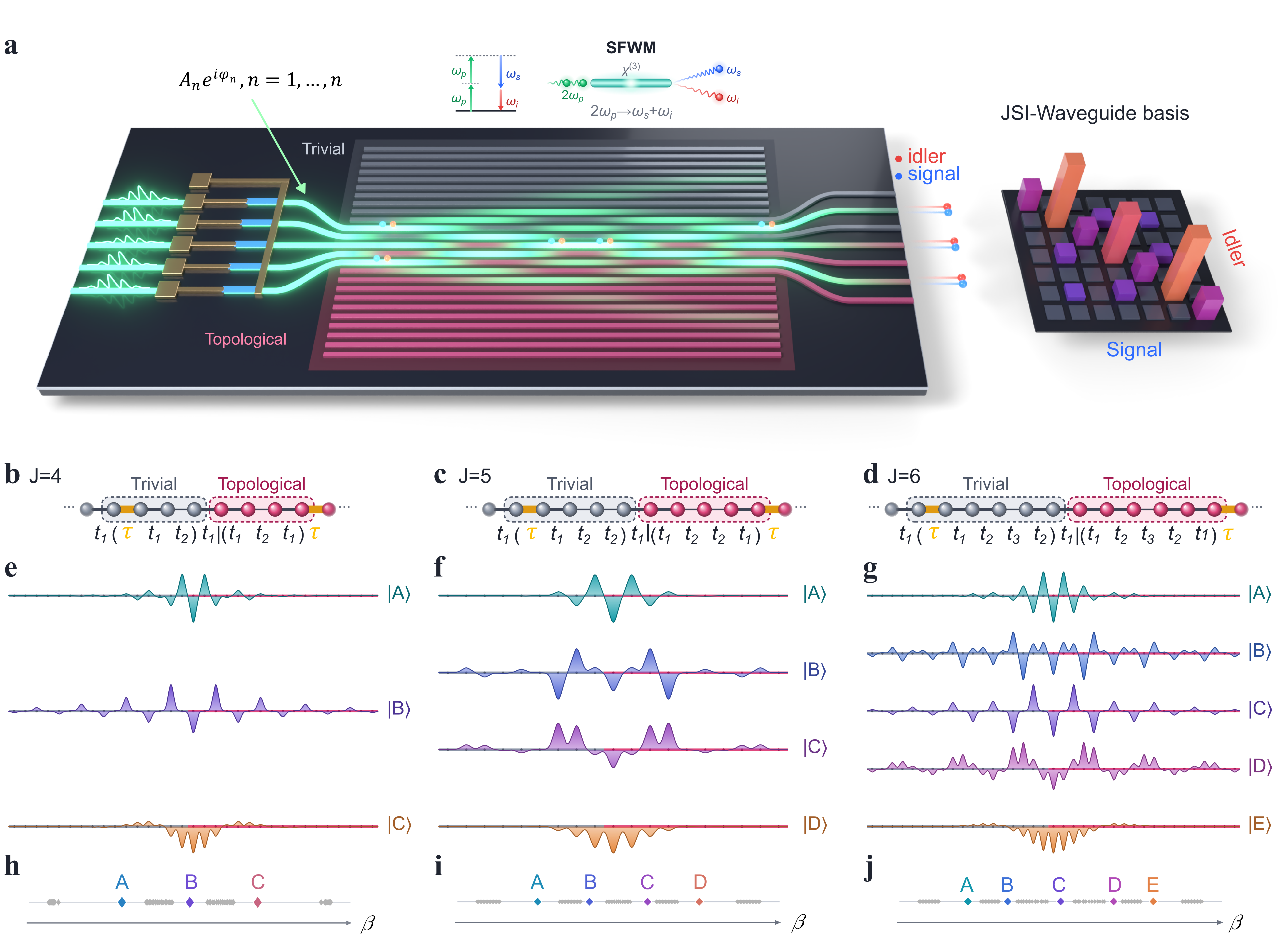}
\caption{\textbf{Programmable silicon topological superlattice for high-dimensional biphoton generation.}
\textbf{a,} Schematic of the silicon photonic platform. Five programmable pump inputs address the central interface region, seven output waveguides collect the generated signal--idler correlations. Independent amplitude and phase control of the pump inputs prepares coherent excitations of the localized interface-mode manifold. Through SFWM, the programmed pump selects the nonlinear pathways that populate the output biphoton state.
\textbf{b--d,} Coupling sequences of the $J=4-6$ superlattices, respectively, formed by joining domains with distinct hopping patterns leading to distinct topologies. \textbf{e--g,} Topological modal profiles for the $J=4-6$ superlattices, respectively, showing the three-, four- and five-mode localized interface manifolds that form the basis for programmable biphoton generation. \textbf{h--j}, Propagation-constant ($\beta$) spectra for $J = 4$--6. Each topological mode (coloured; A--E as in \textbf{e--g}) sits in its own gap between the grey bulk bands, giving $J-1$ topological interface modes for each lattice.}
\label{Fig1}
\end{figure}

\noindent The central piece of this investigation is multiband topological superlattices excited by a coherent multi-port pump [Fig.~\ref{Fig1}a]. The superlattices consist of evanescently coupled silicon photonic waveguides with distinct hopping sequences arranged into periodically repeated unit cells with an increasing number of waveguides \(J=4, 5\) and 6 [Figs.~\ref{Fig1}b,c, and d]. As established in Ref. \cite{PhysRevA.98.043838}, a superlattice of order \(J\) hosts \(J-1\) topological bandgaps, as long as the intracell hopping strengths $t_1$, $t_2$ and $t_3$ preserve inversion symmetry and the intercell hopping strength $\tau$ remains larger than a critical value at which the smallest topological bandgap closes. Topologically protected interface states can be induced in all of these topological band gaps when a topological lattice with these characteristics is interfaced with a trivial (non-topological) one, as depicted in Figs.~\ref{Fig1}b,c, and d for superlattices with 4, 5 and 6 waveguides per unit cell respectively. These topological interface states are co-localized at the trivial-topological transition but show diverse modal profiles [Figs.~\ref{Fig1}e--g] and propagation constants $\beta$ [Figs.~\ref{Fig1}h--j]. As evidenced by these figures, the accessible modal space is increased with \(J\), enlarging the number of protected pump and biphoton channels available for quantum state synthesis. 

The chip is accessed through an array of grating couplers connected to five programmable pump inputs, seven signal--idler collection outputs, and an auxiliary waveguide for alignment and coupling calibration (see Methods). The pump inputs spans the interface region, allowing the launched field to be prepared as a coherent superposition of selected waveguide excitations. By tuning their relative amplitudes and phases, the pump profile can be matched to individual topological interface modes or to optimized superpositions of such modes (see Supplementary Information S1). Because the low-gain SFWM regime depends on the squared pump field, this coherent input state determines the nonlinear overlap that seeds the generated signal--idler wavefunction.

The underlying physics for this can be explained as follows. The pump amplitude in each waveguide $n$ as a function of the propagation distance can be expanded on the basis of the lattice eigenmodes as
\begin{equation}
A_n(z)=\sum_jc_j(z)f_{j,n}^p, 
\label{eq1}
\end{equation}
where $f_{j,n}^p$ is the amplitude of the pump $j-$th mode in the $n-$th waveguide and $c_j(z)$ is the coefficient of that mode in the coherent pump superposition, which we control via our programmable pump inputs. These weights do not stay constant upon propagation in the superlattice, but evolve as
\begin{equation}
c_j(z)=c_j(0)e^{-i\beta_j^pz},
\label{eq2}
\end{equation}
where $\beta_j^p$ is the propagation constant of the $j-$th mode. The biphoton state generated from the nonlinear interaction of this pump with the medium can be expressed as
\begin{equation}
\psi(z)=\sum_{l,m}\phi_{l,m}(z)b_l^\dagger b_m^\dagger, 
\label{eq3}
\end{equation}
where $\phi_{l,m}$ is the probability amplitude that a signal photon occupies mode $l$ and the idler photon occupies mode $m$. The evolution of this probability follows from the equation
\begin{equation}
i\frac{\partial\phi_{l,m}}{\partial z}=(\beta_l^s +\beta_m ^i)\phi_{l,m}+\sum_{j,k,l,m}\Gamma_{jklm}c_j c_k, 
\label{eq4}
\end{equation}
with the term $(\beta_l^s +\beta_m ^i)\phi_{l,m}$ accounting for free propagation while $\sum_{j,k,l,m}\Gamma_{jklm}c_j c_k$ is the nonlinear source term, which depends on the pump coefficients. The nonlinear overlap, introduced originally in Ref.\cite{Doyle2022biphoton}, 
\begin{equation}
\Gamma_{jklm}=\gamma\sum_{jklm}f_j^pf_k^pf_l^{s*}f_m^{i*} , 
\label{eq5}
\end{equation}
measures the overlap of pump mode profiles $j$ and $k$ with signal mode $l$ and idler mode $m$. The value of $\Gamma_{jklm}$ can be understood as the selection rule for SFWM: larger nonlinear overlaps lead to efficient photon pair generation in a given combination of modes ${j,k,l,m}$ and if the overlap becomes zero, then SFWM simply cannot not occur between those modes. Therefore, by controlling the pump modal coefficients $c_j$ and $c_k$ we can enhance or decrease the efficiency with which two pump photons in modes $j$ and $k$ generate a signal-idler photon pair in the modes ${l,m}$, and in this way we gain full control over the preparation of the resulting output entangled state (see Supplementary Information S2).
Our approach separates two functions that are normally fixed together in integrated nonlinear sources. The silicon photonics superlattice supplies a disorder-resilient modal resource, whereas the coherent pump determines how that resource is populated during spontaneous four-wave mixing. The result is a fixed topological chip that can be reconfigured at the source level, enabling control over the modal pathway, spatial correlation structure, and effective dimensionality of the generated biphoton state.\\

\noindent\textbf{Selective topological mode excitation}\\

\begin{figure}[H]
    \centering
    \includegraphics[width=1\textwidth]{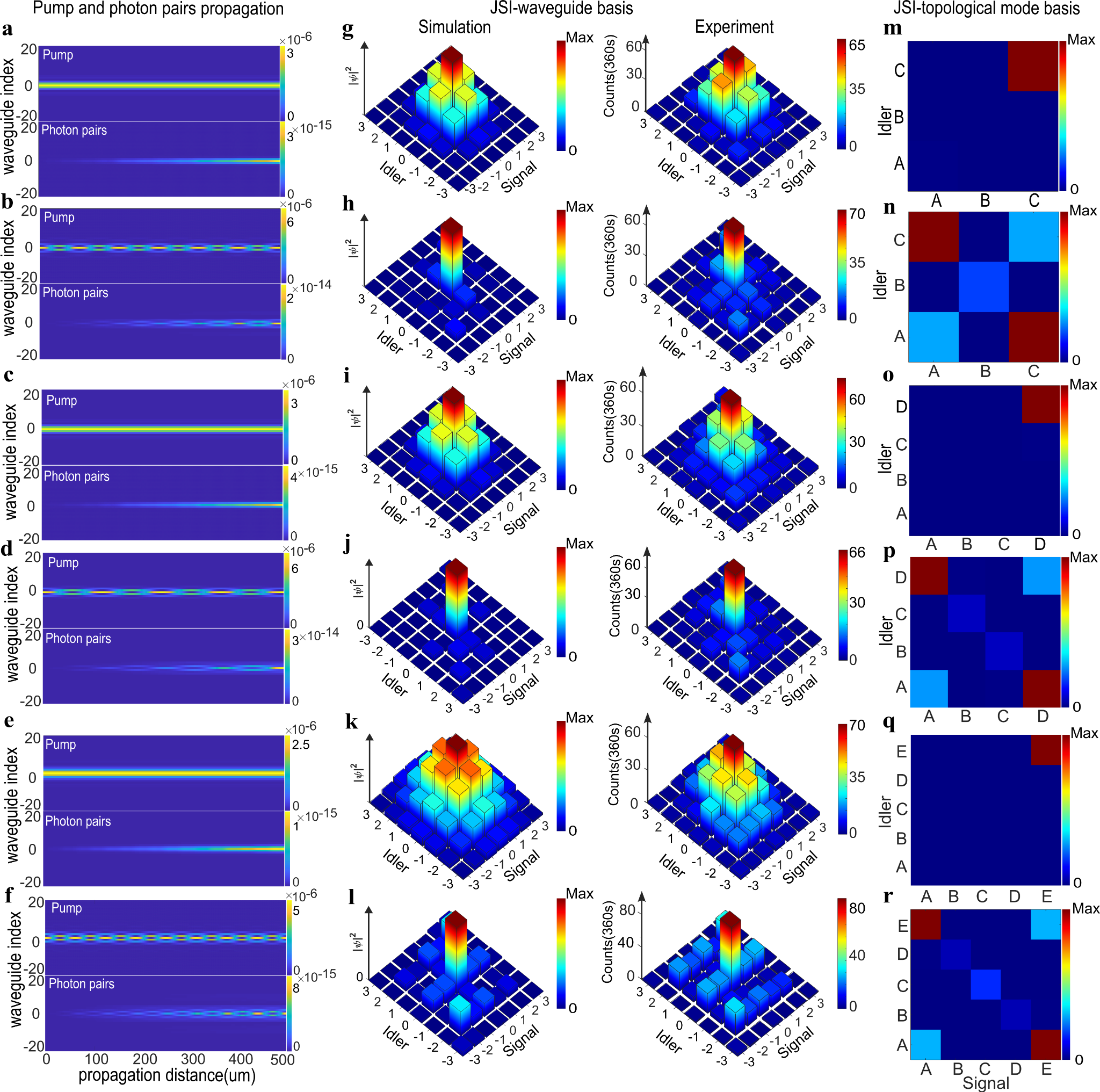}
\caption{\textbf{Selective topological mode excitation.}
\textbf{a,b,} Pump and photon-pair propagation simulations in the \(J=4\) lattice for excitation profiles matching topological pump mode $|C\rangle$ and a coherent superposition $|A\rangle+|C\rangle$, respectively.
\textbf{c,d,} Same in the \(J=5\) lattice with excitations matching topological pump mode $|D\rangle$ and $|A\rangle+|D\rangle$, respectively.
\textbf{e,f,} Same in the \(J=6\) lattice with excitations matching $|E\rangle$ and $|A\rangle+|E\rangle$, respectively.
\textbf{g--l,} Simulated and measured output JSI in the waveguide basis $-$ shown in the seven central waveguides, indexed from \(-3\) to \(+3\) with waveguide 0 at the center of the topological transition $-$ for the six pump configurations in \textbf{a--f}. The simulations show the normalized intensity probability \(|\psi|^2\) and the experiments show the coincidence events over \(360\,\mathrm{s}\).
\textbf{m--r,} Corresponding output projected BEP matrices.} 

    \label{Fig2}
\end{figure}

\noindent Pump-defined biphoton state generation is first demonstrated using two classes of programmed excitation: individual topological pump eigenmodes and coherent superpositions (sums) of two topological pump eigenmodes.  Figure~\ref{Fig2} presents the response of superlattices of increasing number of waveguides per unit cell to these two classes of excitation. 
For individual eigenmode excitation, the propagation simulations confirm that both the pump and the photon pairs generated via SFWM along the lattice remain confined to a single topological mode [Fig.~\ref{Fig2}a,c,e] (see Supplementary Information S3). The quantum states at the output of the lattice are experimentally characterized via the JSI in the waveguide basis, using a combination of spectral filtering, single-photon detection, and time-correlation inference electronics. Coincidence-to-accidental ratios (CARs) vary across output channels but are predominantly in the 20–30 range for populated channels, with typical \(1\sigma\) uncertainties of 1–2 (see Supplementary Information S4). The measured biphoton correlation maps agree well with the output of the quantum propagation simulations [Fig.~\ref{Fig2}g,i,k]. These states are measured in the lattice waveguide basis, but their projection onto the biphoton eigenmode basis $-$ the biphoton eigenmode population (BEP) $-$ reveals that selective topological mode pump excitation translates into selective topological mode biphoton population [Fig.~\ref{Fig2}m,o,q] (See Supplementary Information S5 and S6) 

Coherent superpositions of two topological pump modes lead to a beating pattern of such two modes upon propagation, given that they have different propagation constants [Fig.~\ref{Fig2}b,d,f]. The photon pair propagation simulations also show a beating pattern that looks slightly different from the pump one, indicating that other modes may be involved in the nonlinear interaction. The output states on the waveguide basis [Fig.~\ref{Fig2}n,p,r] and their projection onto the eigenmode basis [Fig.~\ref{Fig2}h,j,l] confirm this hypothesis: coherent superpositions of two pump modes lead to enhanced biphoton population of those two modes, but also non-negligible biphoton population of modes that were not populated by the pump. 


\noindent\textbf{Arbitrary quantum state generation}\\

Having established mode selectivity, we now focus on correlation steering by pump optimization. As a first target, we optimize the pump to enhance the population of a selected signal–idler eigenmode block relative to the others. The goal is not to suppress all other modes, but to show that we can steer where the biphoton state resides within the signal–idler eigenmode space.

\begin{figure}[H]
    \centering
    \includegraphics[width=1\textwidth]{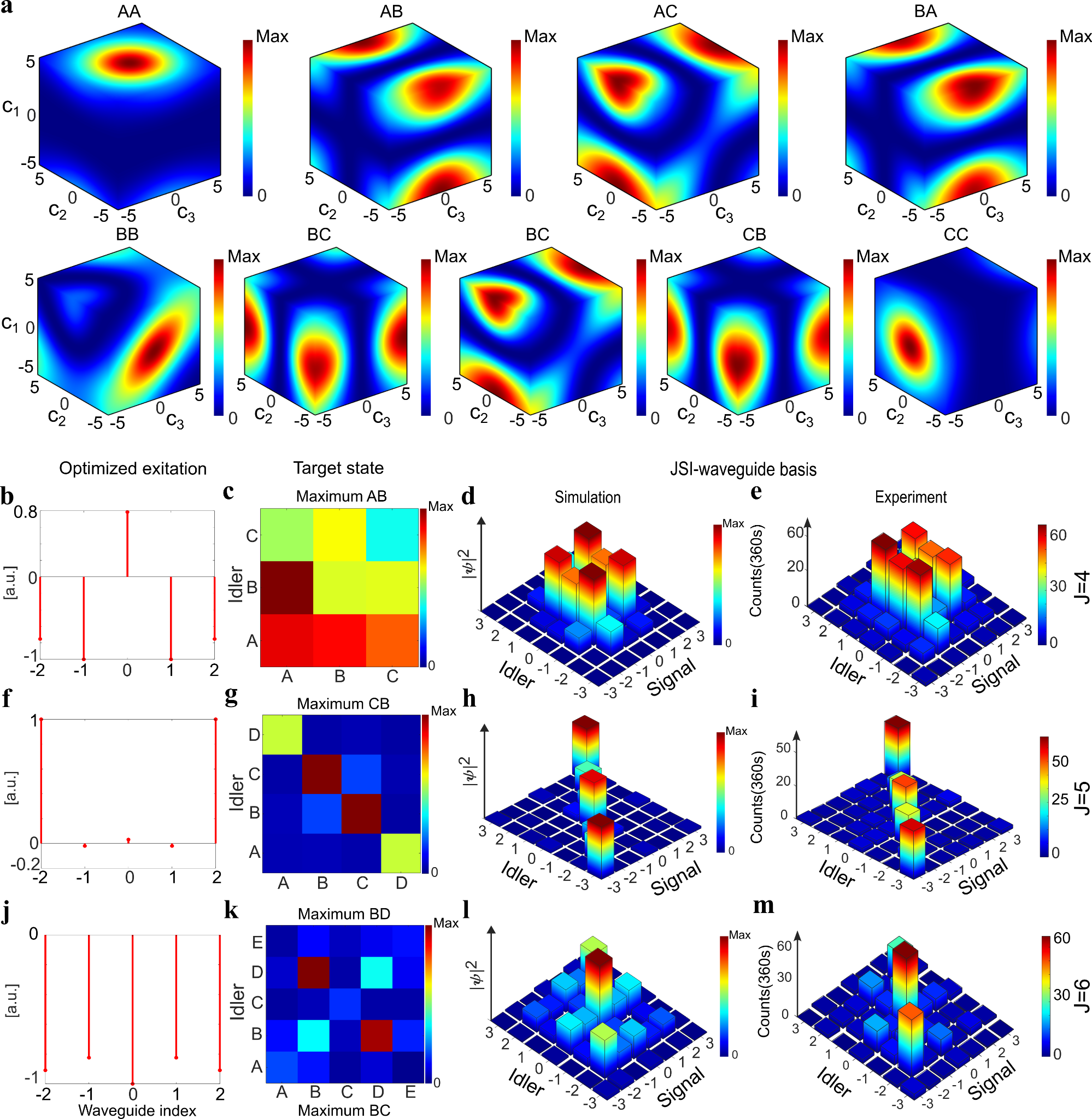}
\caption{\textbf{Optimization-driven routing of biphoton correlations in topological-mode space.}
\textbf{a,} Pump-coefficient control landscape of the $J=4$ lattice. Biphoton eigenmode populations of signal--idler blocks versus the coefficients $(c_1,c_2,c_3)$ of the coherent superposition of the three topological modes at fixed total power. \textbf{b,} Optimized pump excitation required to achieve a target quantum state with a local maximum probability at the off-diagonal block \(AB\).\textbf{c,} Output BEP for such target quantum state at the output of the \(J=4\). \textbf{d,e} Simulated and measured output quantum states on the waveguide basis. \textbf{f-i, j-m} Same as b-e for the target quantum states shown in e and i sampling a variety of entanglement shapes with maximum probability at off-diagonal and on-diagonal blocks.} 

    \label{Fig3}
\end{figure}

To illustrate how coherent pump control can preferentially weight chosen signal–idler correlations while preserving a multimode quantum state, Fig.~\ref{Fig3}a shows the weight of each individual block of the output state in the topological mode basis ($|AA\rangle, |AB\rangle,...|CC\rangle)$) for the  $J=4$ lattice, as the coefficients of the coherent pump superposition $(c_1,c_2,c_3$ in Eq. \ref{eq1}) are scanned . Whereas a single-mode pump fixes the generated state through its spatial overlap with the signal--idler modes, here three real coefficients reshape the entire $3\times3$ biphoton population matrix on a disorder-protected modal basis. Targeted state generation thereby reduces to a low-dimensional inverse-design problem, which we solve by pump optimization (Supplementary Information S8). 

This capability is experimentally demonstrated in the $J=4$ superlattice by choosing an optimized combination of coefficients $[c_1=-2.6,c_2=-3.4,c_3=2.1]$ to maximize $|AB\rangle$. The resultant excitation $A_n(0)$ $-$ obtained using those coefficients in see Eq. (1) $-$ is shown in Fig.~\ref{Fig3}b. Notice that we have truncated the excitation to the five center waveguides available as experimental inputs, which we have found to be sufficient to achieve the desired output quantum states (see Supplementary Information S3). Our propagation simulations using such excitation lead to the target state in the topological modal basis and in the waveguide basis shown in Fig.~\ref{Fig3}c and d, respectively. The measured JSI in the waveguide basis under the selected pump excitation agrees well with the simulated target state Fig.~\ref{Fig3}e, experimentally verifying the targeted correlation steering. We performed analogous experimental demonstrations of controlled correlation steering in the $J=5$ and $6$ superlattices, finding coherent pump excitations of four [Fig.~\ref{Fig3}f] and five topological modes [Fig.~\ref{Fig3}j] to maximize $|CB\rangle$ [Fig.~\ref{Fig3}g] and $|BD\rangle$ [Fig.~\ref{Fig3}k], respectively. In these cases, the measured quantum states [Figs.~\ref{Fig3}i,m] also agreed well with the simulations [Figs.~\ref{Fig3}h,l], and the targeted state was experimentally reproduced with a mean fidelity of \(\approx 94\%\) (see Supplementary Information S9).

Together, Figs.~\ref{Fig2} and \ref{Fig3} establish a hierarchy of control over topological photon-pair generation. Single-mode pumping selects individual biphoton channels; coherent two-mode pumping redistributes amplitude within the protected manifold; and optimized multimode pumping routes correlations into prescribed eigenmode blocks. This hierarchy is the operational basis of the platform: a fixed silicon superlattice can generate different topological biphoton states solely by reconfiguring the coherent pump see Movies 1, 2 and 3.

\noindent\textbf{Programmable entanglement dimensionality}\\

Programmable access to individual topological mode pairs gives control over the microscopic pathways of photon-pair generation. A more demanding goal is to fix a collective property of the quantum state before it is produced. In our experiments, we choose this collective property to be the dimensionality of the entangled state, quantified by the Schmidt number (K). In conventional integrated sources this dimensionality emerges from the fixed device geometry and is evaluated only after generation; here, we can design it via our controlled coherent pump. To do this, we solve the following inverse problem: finding the modal coefficients in the coherent pump superposition $c_j$ [Eq. \ref{eq1}] that produce biphoton states of prescribed dimensionality. The search is carried out in the topological pump mode basis, but each selected solution is verified by direct propagation of the coupled pump and two-photon dynamics in the waveguide basis, so that the designed states follow from the full nonlinear evolution of the superlattice rather than from a reduced modal approximation (see Supplementary Information S10).

\begin{figure*}[t]
  \centering
  \includegraphics[width=0.8\textwidth]{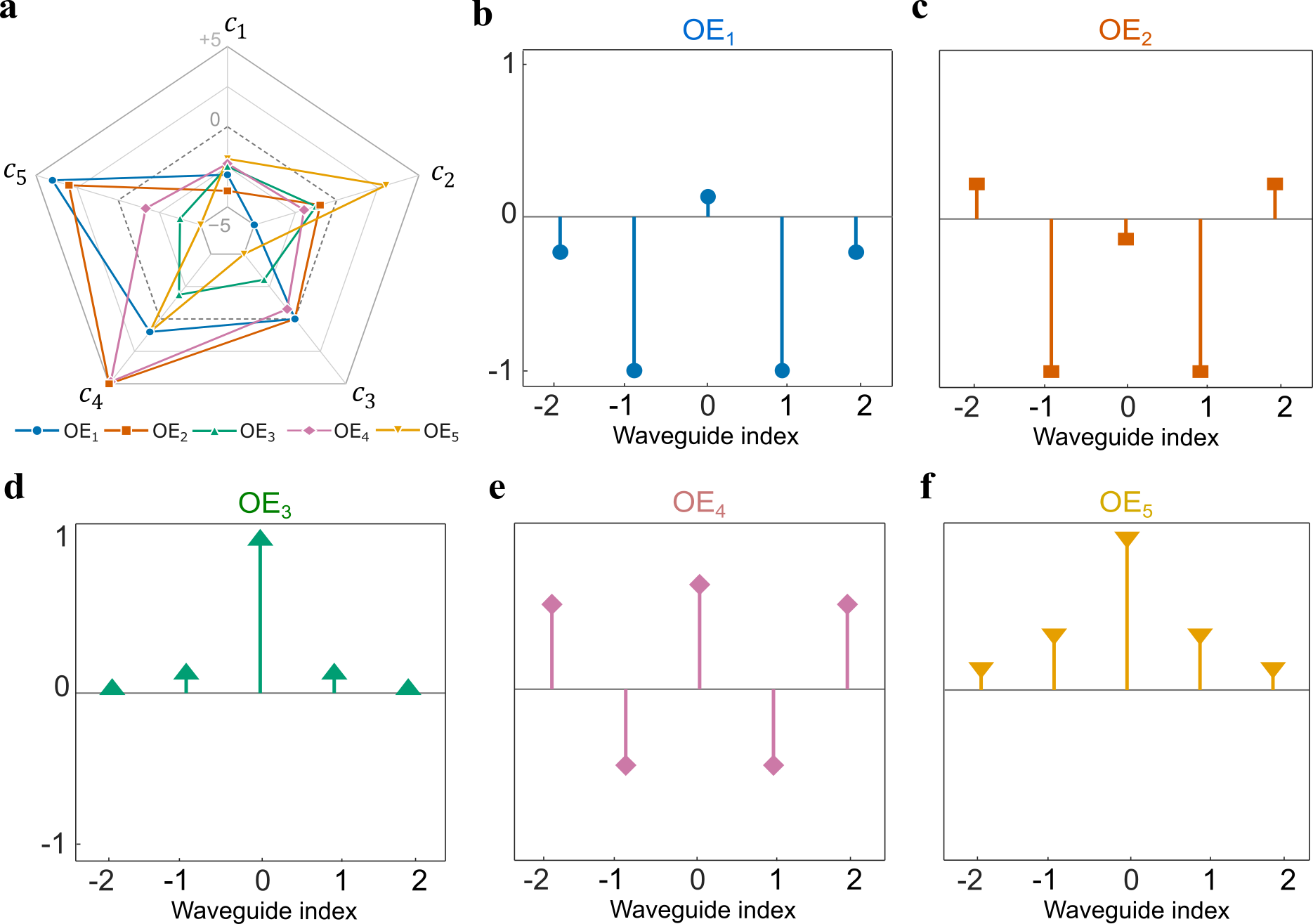}
  \caption{\textbf{Optimized excitation profiles for a targeted quantum state dimensionality ($K_{target}=5$).}
  \textbf{a},~Inverse-designed pump-control vectors $\mathbf{c}=(c_1,c_2,c_3,c_4,c_5)$ of the
  $J=6$ superlattice for the five solutions A--E with $K_{\mathrm{target}}=5$.
  \textbf{b}--\textbf{f},~Corresponding normalized launch profiles; colors as in \textbf{a}.}
  \label{Fig4}
\end{figure*}

Not surprisingly, a targeted dimensionality can be achieved with a plethora of different coherent superpositions. 
Figure~\ref{Fig4}a shows five different combinations for the five modal pump coefficients $c_1$, $c_2$, $c_3$, $c_4$, and $c_5$ to achieve a targeted dimensionality $K_{\mathrm{target}}=5$ in a $J=6$ superlattice. The five optimized excitation profiles resulting from these combinations of coefficients are shown in Figs.~\ref{Fig4}b-f. All these profiles were verified to reproduce the targeted dimensionality through direct numerical simulations. 


\begin{figure}[H]
  \centering
  \includegraphics[width=\textwidth]{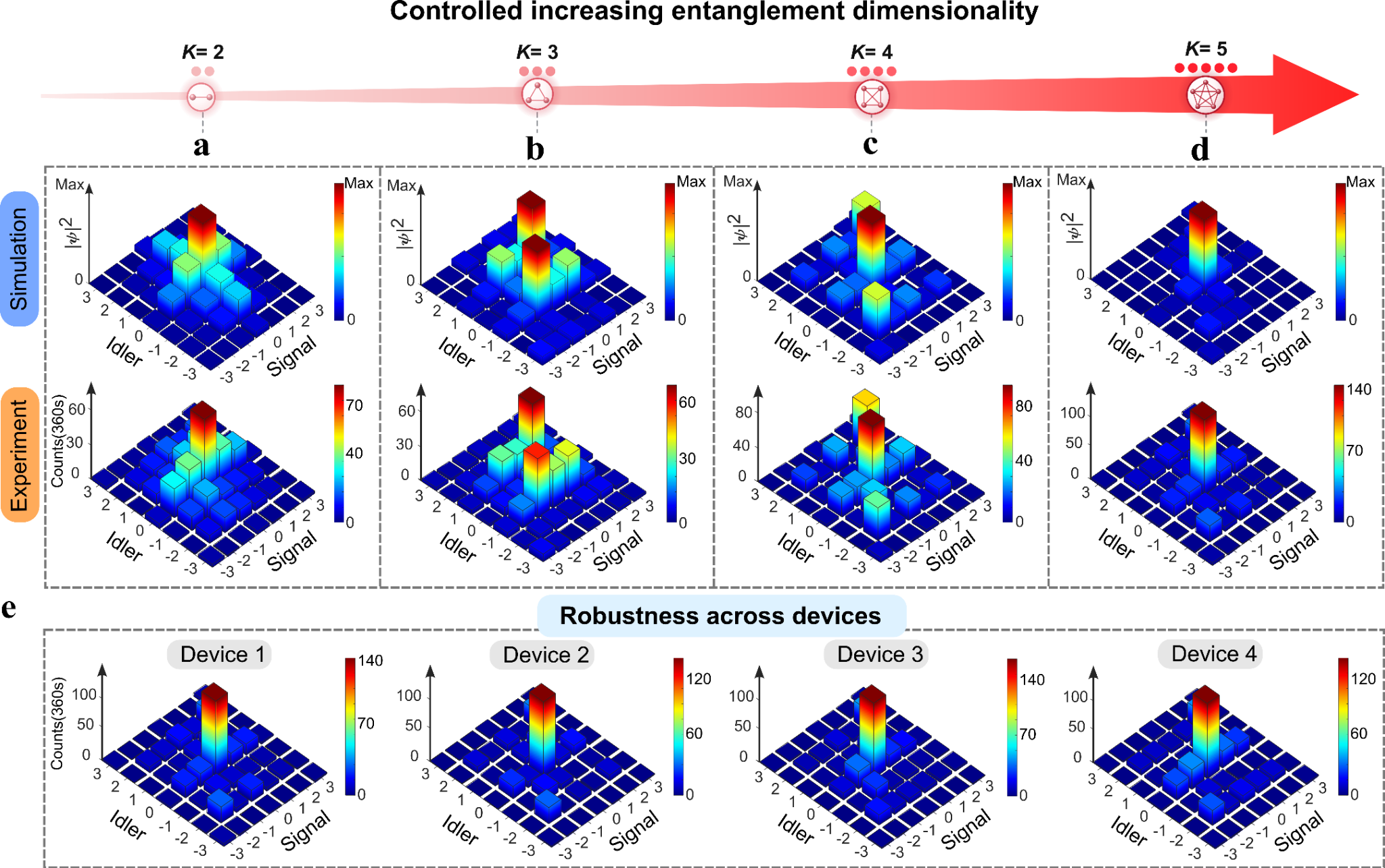}
  \caption{\textbf{Inverse-designed generation of high-dimensional topological biphoton entanglement.}
  Spatially entangled signal--idler photon pairs are generated by spontaneous four-wave mixing in silicon topological superlattices. For each superlattice order \(J\), the input pump is prepared as a coherent superposition of localized interface modes whose amplitudes are inverse-designed to synthesize a prescribed Schmidt number \(K_{\mathrm{target}}\). Coincidences are measured between the seven central output waveguides, indexed from \(-3\) to \(+3\), using superconducting-nanowire single-photon detectors over a \(360\,\mathrm{s}\) integration time.
  \textbf{a--d,} Simulated and measured joint signal--idler intensity distributions for \(J=6\), targeting \(K_{\mathrm{target}}=2\), \(3\), \(4\), and \(5\), respectively. Increasing the target Schmidt number produces progressively richer spatial-correlation patterns, consistent with higher-dimensional biphoton entanglement.
  \textbf{e,} Reproducibility of the \(K_{\mathrm{target}}=5\) state across four nominally identical fabricated devices. Preservation of the dominant correlation structure demonstrates the robustness of the programmed biphoton state against fabrication-induced variations.}
  \label{Fig5}
\end{figure}

We experimentally demonstrate programmable control over the dimensionality of the output states in the \(J=6\) superlattice, targeting increasing Schmidt numbers \(K_{\mathrm{target}}=2\) to \(5\) [Figs.~\ref{Fig5}a--d]. In all cases, we chose one of the multitude of pump profiles that would render the target. To verify the robustness of this method to achieve programmable state dimensionalities we input the pump profile used to achieve \(K_{\mathrm{target}}=5\) into four other different devices in the same chip and measured the JSI [Fig.~\ref{Fig5}e]. 

\begin{figure}[H]
  \centering
  \includegraphics[width=\textwidth]{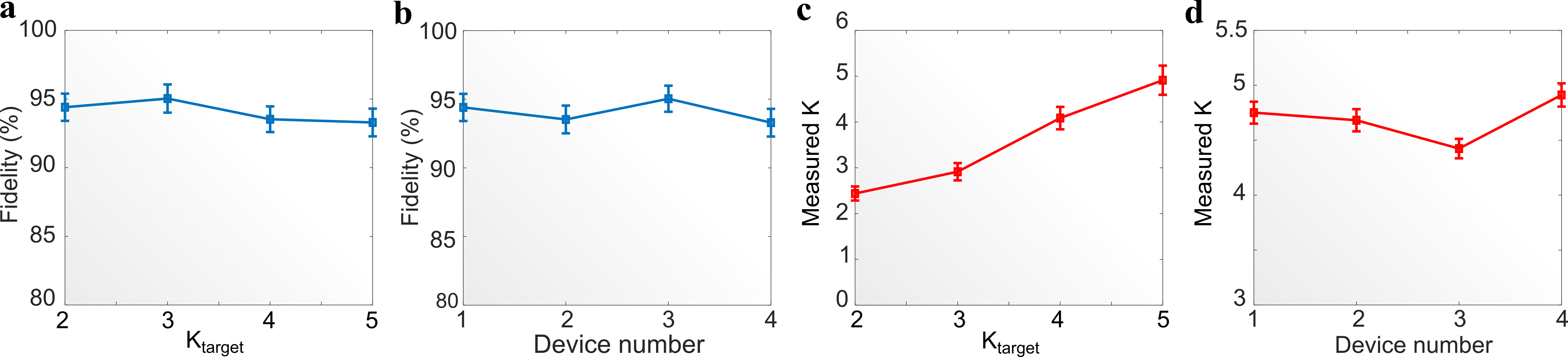}
  \caption{\textbf{Measured fidelity and entanglement dimensionality in the $J=6$ superlattice.} \textbf{a,} Fidelity of the measured joint spatial intensities for increasing target Schmidt numbers. \textbf{b,} Fidelity of the measured joint spatial intensities accross devices for the target dimensionality $K_{\mathrm{target}}=5$. \textbf{c,} Lower bound on the Schmidt number $K$ for the measured states versus $K_{\mathrm{target}}$. \textbf{d,} Lower bound on the Schmidt number across devices for $K_{\mathrm{target}}=5$.}
  \label{Fig6}
\end{figure}

As evidenced by Figs.~\ref{Fig5}a--d, the measured JSIs in the waveguide basis for increasing dimensionality agree well with those simulated. In Fig.~\ref{Fig6}a we quantify this agreement using the fidelity metric and obtain values $>94\%$. This high fidelity is maintained across measurements in different devices, as shown in Fig.~\ref{Fig6}b. Finally, a lower bound for the Schmidt number of the measured quantum states \cite{Eckstein2014} as a function of the targeted Schmidt number is shown in Fig.~\ref{Fig6}c, evidencing that the measured states have the desired dimensionality. Disorder resilience is quantified in Figure ~\ref{Fig6}d, which shows the lower bound for the Schmidt number of the measured states across devices shown in Fig.~\ref{Fig5}e, targeting \(K_{\mathrm{target}}=5\).


\section*{Conclusion}

We have demonstrated a programmable silicon topological superlattice in which high-dimensional biphoton entanglement is engineered within a protected multimode interface manifold. By coherently shaping the pump across localized topological modes, spontaneous four-wave mixing is converted from a geometry-defined emission process into an inverse-designable operation on the biphoton wavefunction. This enables control over the generated state at multiple levels, including its signal--idler correlations, topological modal pathway and effective Schmidt dimensionality.

The key advance is the separation between modal protection and nonlinear state synthesis. The multiband superlattice provides a disorder-resilient Hilbert space that expands with superlattice order, while the pump determines how this space is populated. As a result, distinct high-dimensional entangled states can be generated on a fixed silicon device without changing the fabricated structure. Their reproducibility across nominally identical electron-beam-lithography-fabricated chips further shows robustness to realistic nanofabrication disorder.

Beyond demonstrating programmable control of topological entanglement, the results presented here establish a key ingredient for exploiting topological photonic states as computational resources. Most photonic quantum information protocols require the preparation of specific entangled input states rather than generic photon-pair generation. Directly tailoring the dimensionality and modal distribution of the generated state can reduce the quantum circuit depth required for state synthesis, facilitate the initialization of high-dimensional quantum registers, and provide application-specific resource states for quantum communication, networking and measurement-based photonic quantum computing. As interest grows in qudit-based photonic processors, the ability to engineer topological entangled states with prescribed quantum-state characteristics offers a versatile route toward programmable quantum photonic architectures.


\section*{Methods}
\textbf{Design and fabrication methods}\\
The layouts were assembled from foundry-qualified parameterized cells,
verified against process design rules, and exported to GDSII for fabrication. The devices were fabricated on silicon-on-insulator (SOI) wafers with a
\SI{220}{\nano\metre} silicon device layer at the Applied Nanotools
NanoSOI foundry. Waveguide patterns were defined by high-resolution
electron-beam lithography, transferred into silicon by inductively coupled
plasma etching, and encapsulated with a \(\sim\SI{2.2}{\micro\metre}\)
\(\mathrm{SiO_2}\) top cladding. The large index contrast between silicon
and silica provides strong optical confinement, while the oxide cladding
improves environmental stability, reduces surface-related scattering, and
supports reproducible long-duration quantum-optics measurements.

\noindent\textbf{Implementation of programmable excitation}

\noindent Programmable control is provided by tunable attenuators and thermo-optic
phase shifters integrated on the five input waveguides. Metal microheaters
placed above straight waveguide sections locally tune the silicon effective
index, enabling independent phase control \(\phi_j\) at each input. By
programming the input phases, the interference landscape of the multiport
lattice can be reconfigured without changing the physical device.

\noindent\textbf{Coupling mechanism}

\noindent Optical input and output coupling is achieved using fully etched,
TM-polarized focusing grating couplers designed for an \ang{8} fiber launch
angle and operation near \SI{1550}{\nano\metre}. The couplers exhibit an
insertion loss of approximately \SI{6.3}{\decibel} per coupler. Adiabatic
tapers are included before the first bends to suppress radiation loss. The
grating couplers are arranged for a 16-channel fiber array with
\SI{250}{\micro\metre} pitch, comprising five inputs, seven outputs, and two
alignment/reference channels (Fig.~\ref{Fig1}).

\backmatter

\bmhead{Supplementary information}

Supplementary information is available for this paper, including Supplementary Notes S1--S10, Supplementary Figs. S1--S22, Supplementary Tables S1--S3. Supplementary Movies 1--3 illustrate the programmed pump dynamics, photon-pair generation, and modal evolution associated with representative topological excitation protocols.

\bmhead{Acknowledgements}

This work is sponsored by the National Science Foundation (NSF) grant number 2328993 and by the Army Research Office under Grant Number W911NF-26-1-A206. The views and
conclusions contained in this document are those of the authors and should not be interpreted as representing the official policies, either expressed or implied, of the Army Research Office or the U.S. Government. The U.S. Government is authorized to reproduce and distribute reprints for
Government purposes notwithstanding any copyright notation herein.

\section*{Declarations}

The data that support the findings of this study are available from the corresponding author on reasonable request.

\clearpage
\setcounter{page}{1}                          
\renewcommand{\thepage}{S\arabic{page}}
\setcounter{figure}{0}\renewcommand{\thefigure}{S\arabic{figure}}
\setcounter{table}{0}\renewcommand{\thetable}{S\arabic{table}}
\setcounter{equation}{0}\renewcommand{\theequation}{S\arabic{equation}}

\begin{center}
  {\Large\bfseries Supplementary information for}\\[12pt]
  {\LARGE\bfseries Programmable Preparation of Topological Entangled States}\\[16pt]
\end{center}

\tableofcontents
\clearpage

\suppheading{S1. Classical simulation and measurements}

\begin{figure}[H]
    \centering
    \includegraphics[width=1\textwidth]{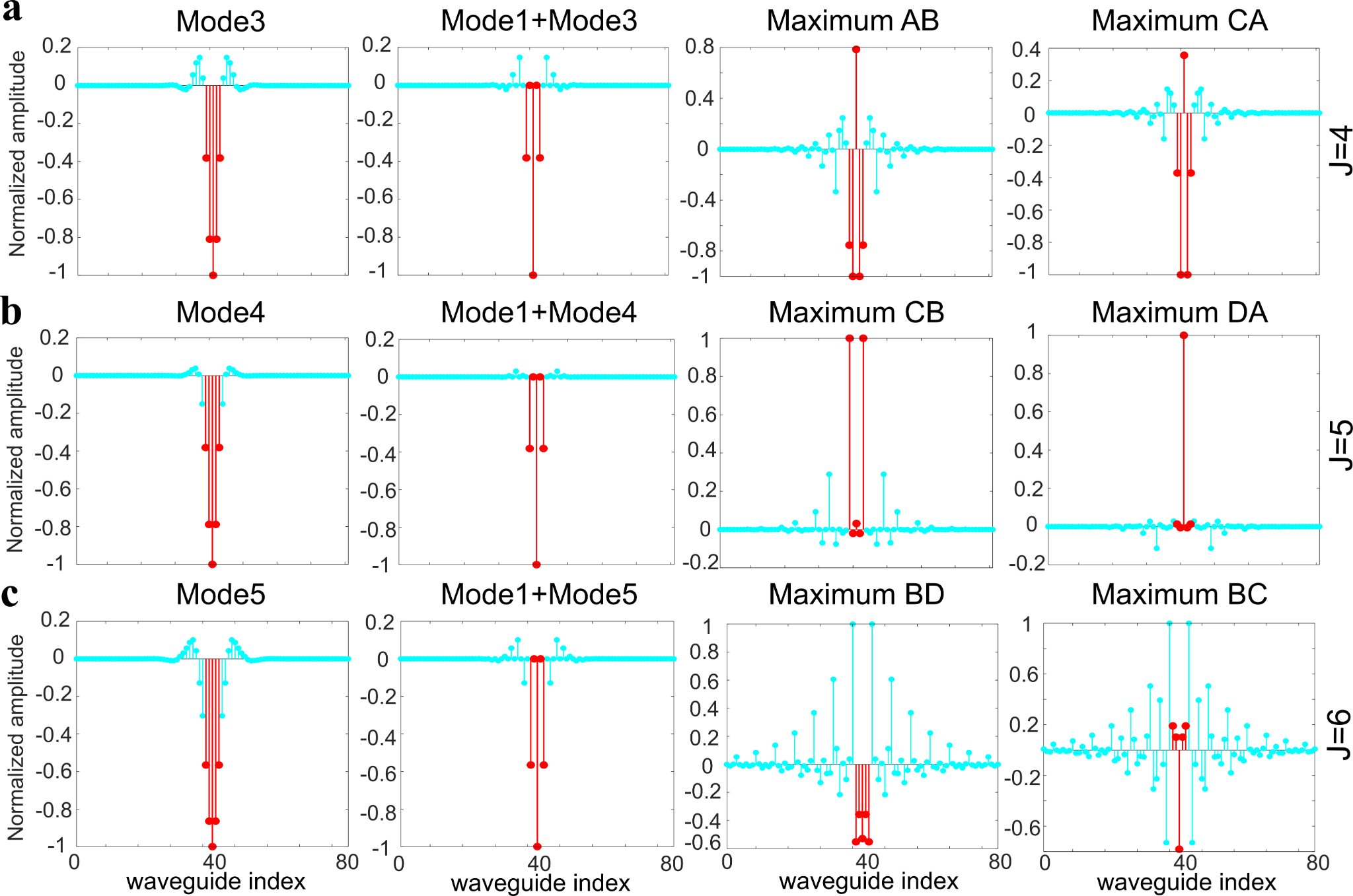}
\caption{\textbf{Programmed topological pump profiles and experimentally addressed input branches.}
Calculated pump-field amplitudes for representative programmed excitations in the \(J=4\), \(J=5\) and \(J=6\) topological superlattices. 
Columns show representative input states, including single localized topological-mode excitation, coherent two-mode excitation and optimized pump profiles designed to enhance selected biphoton eigenmode-population blocks. 
Cyan stems denote the full normalized modal amplitude across the waveguide array, whereas red stems indicate the input waveguide branches addressed in the experimental multi-port excitation scheme. 
These branches sample the localized interface-mode manifold and allow the target pump superposition to be synthesized using independently controlled amplitudes and \(0\) or \(\pi\) phase shifts. 
The figure links the calculated topological pump coefficients to the practical on-chip input patterns used for programmable biphoton-state generation.}
\label{amp}
\end{figure}

A key requirement for programmable biphoton generation is that the optimized pump coefficients can be translated into well-defined spatial excitation patterns at the output of the lattice. To verify this connection, we analyze the pump-power distribution in the seven central waveguides after propagation for representative programmed inputs. This waveguide-resolved observable provides a direct diagnostic of the prepared topological modal superposition, since different modal weights produce distinct interference patterns within the interface region. (see Figure~\ref{amp})

\begin{figure}[H]
    \centering
    \includegraphics[width=1\textwidth]{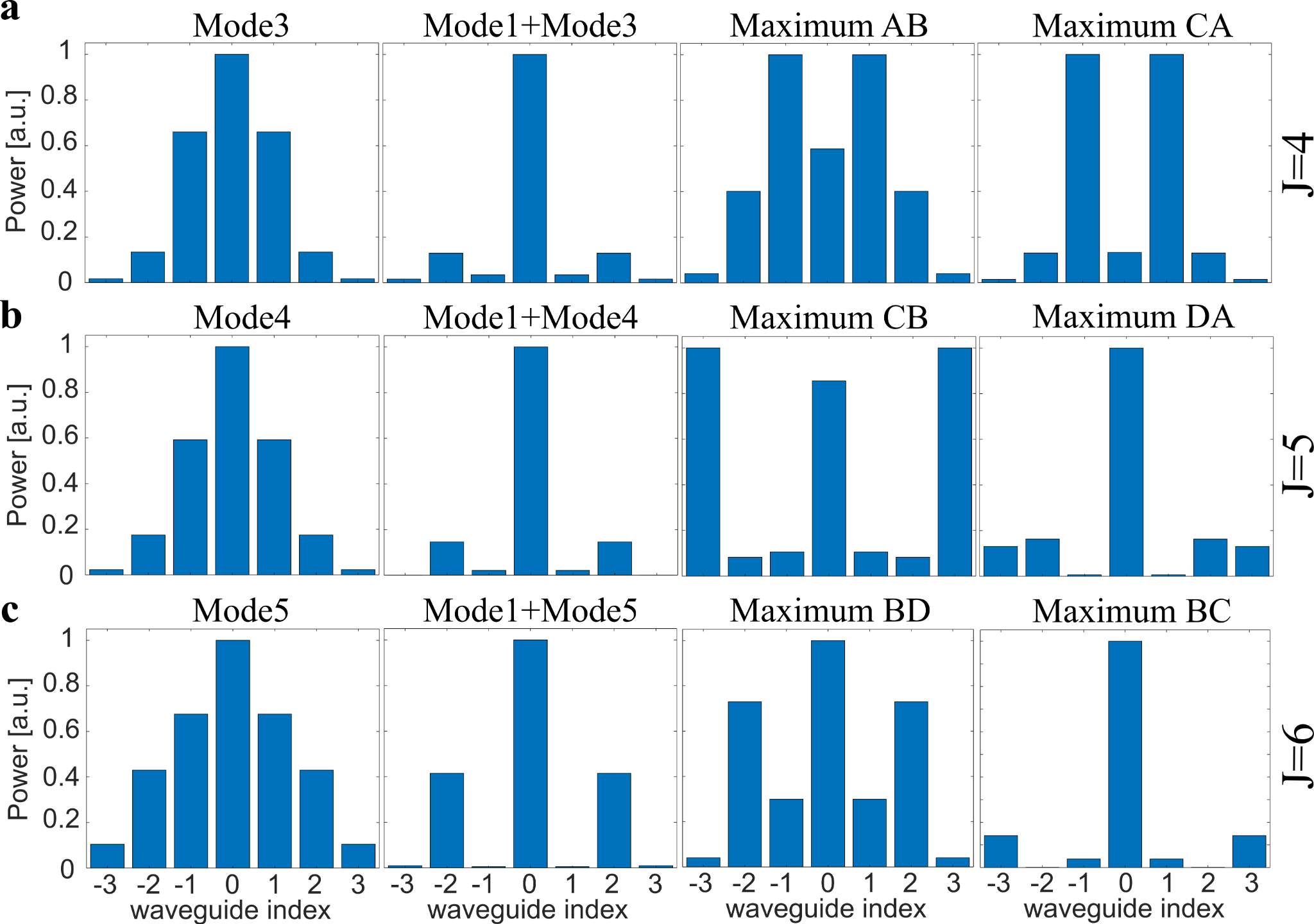}
\caption{\textbf{Simulated pump-output signatures of programmed topological excitations.}
Normalized output pump power calculated in the seven central waveguides at the end of the propagation length for the \(J=4\), \(J=5\) and \(J=6\) topological superlattices. 
Columns compare representative programmed inputs, including single topological-mode excitation, coherent two-mode excitation and optimized multi-port pump states designed to enhance selected biphoton eigenmode-population blocks. 
The different pump superpositions produce distinct spatial fingerprints in the interface region, linking the optimized coefficient vectors to directly observable waveguide-resolved pump distributions. 
These simulations provide a diagnostic for verifying preparation of the desired topological pump superpositions before measuring the corresponding biphoton correlations.}
\label{fig:simulated_output_power}
\end{figure}

Figure~\ref{fig:simulated_output_power} shows the simulated output pump distributions for the $J=4$, $J=5$, and $J=6$ superlattices. In each row, we compare a single-mode excitation, a coherent two-mode superposition, and optimized inputs selected to maximize specific biphoton eigenmode-population blocks. The resulting profiles demonstrate that the optimized coefficient vectors do not correspond to arbitrary input patterns; instead, they generate reproducible spatial signatures determined by the interference of the underlying topological modes. These simulations therefore link the coefficient-space optimization to a measurable pump distribution at the output of the device.

\begin{figure}[H]
    \centering
    \includegraphics[width=1\textwidth]{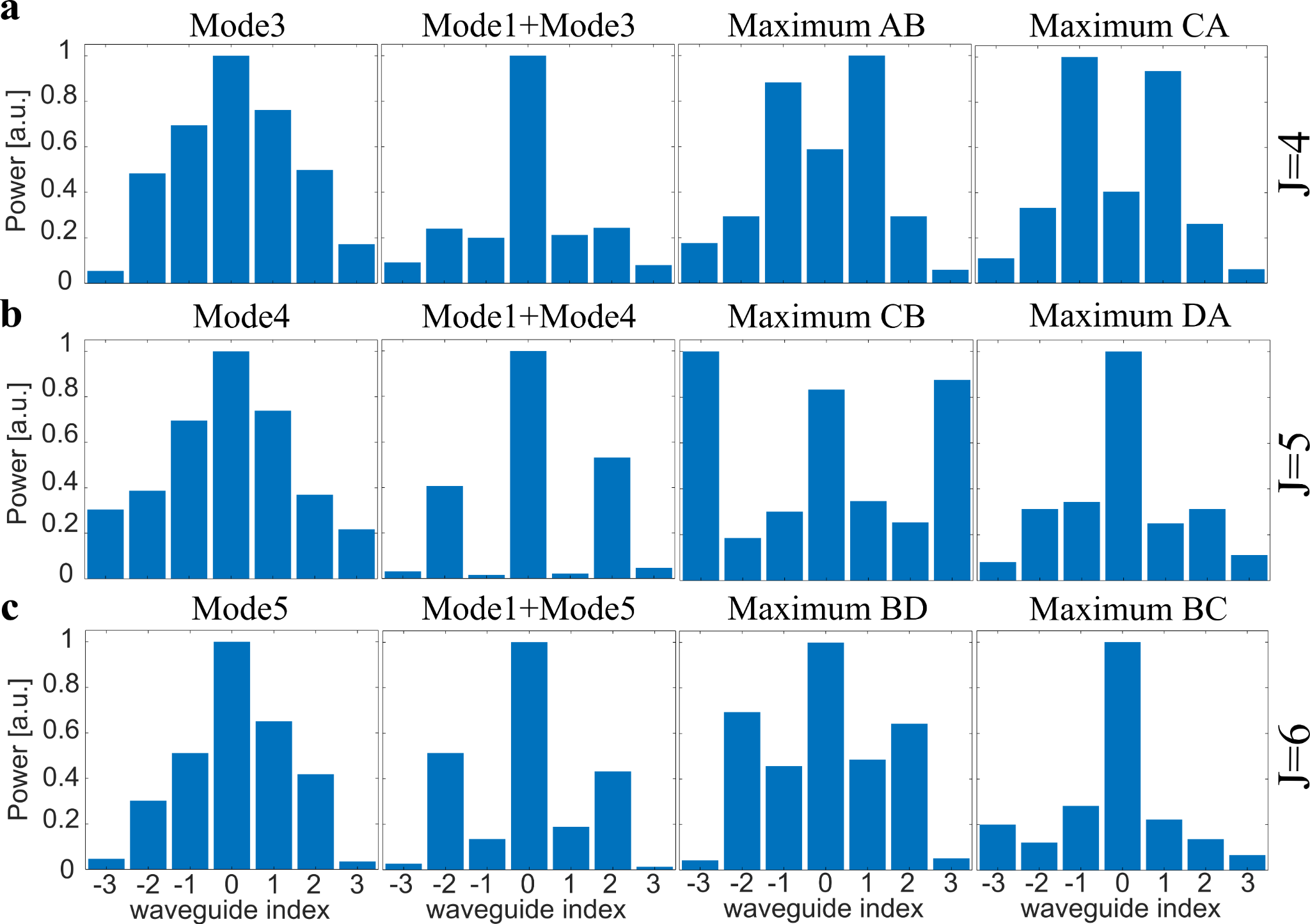}
\caption{\textbf{Measured pump-output signatures of programmed topological excitations.}
Waveguide-resolved output pump power measured after propagation through the \(J=4\), \(J=5\) and \(J=6\) topological superlattices. 
\textbf{a--c,} Measurements for the three-, four- and five-mode interface manifolds, respectively. 
Columns show representative programmed inputs, including single topological-mode excitation, coherent two-mode excitation and optimized multi-port pump states designed to enhance selected biphoton eigenmode-population blocks. 
The horizontal axis denotes the waveguide index relative to the interface centre, and the vertical axis gives the normalized detected pump power. 
The distinct and reproducible output profiles verify the programmed pump preparation and confirm that different optimized input coefficients produce mode-dependent excitation signatures in the interface region. 
These measurements establish the link between coherent input programming and the waveguide-resolved pump fields that drive topological biphoton-state generation.}
\label{fig:experimental_output_power}
\end{figure}

Figure~\ref{fig:experimental_output_power} presents the corresponding experimental measurements. The measured profiles retain the main features predicted by the simulations, confirming that the multi-port excitation scheme can prepare the intended topological modal combinations with good similarity. Differences between the simulated and measured distributions arise from expected experimental imperfections, including coupling imbalance, phase-setting errors, fabrication-induced variations, and residual loss differences among the output channels. Nevertheless, the persistence of distinct mode-dependent output patterns verifies that the optimized input weights are experimentally accessible.

Together, the simulated and measured output-power maps provide an important bridge between the theoretical coefficient-space design and the biphoton-correlation measurements. They show that the programmed pump states used to enhance selected signal--idler eigenmode blocks can be identified through direct waveguide-resolved power readout. This establishes the practical calibration pathway for the experiment: first, the attenuators and phase shifters are tuned until the desired output pump fingerprint is obtained; then, the same excitation condition is used for spontaneous four-wave mixing and coincidence detection. In this way, the output-power distributions serve as an experimentally accessible verification of the pump-state preparation that underlies the programmable control of high-dimensional topological biphoton states.

\suppheading{S2. Theoretical framework: biphoton dynamics in a superlattice }

We consider a discrete superlattice with $N$ sites. The (undepleted) pump field is a site-amplitude vector
$\mathbf{A}(z)\in\mathbb{C}^{N}$, while the biphoton wavefunction is an $N\times N$ matrix
$\psi(z)$ whose element $\psi(m,n)$ is the complex amplitude for detecting the signal photon at site $m$
and the idler photon at site $n$.

Following the undepleted-pump SFWM model, the biphoton obeys the inhomogeneous first-order equation
\begin{equation}
i\,\frac{\partial \psi}{\partial z}
= H_s\,\psi + \psi\,H_i^\dagger + \gamma\,\psi_0\,\mathrm{diag}\!\big(A_1^2(z),\ldots,A_N^2(z)\big),
\label{eq:biphoton_eom}
\end{equation}
where $H_s$ and $H_i$ are the signal and idler tight-binding Hamiltonians, respectively, and the nonlinear
source term is site-local (diagonal) and proportional to the squared pump amplitude.

In the same conventions as the numerical propagation used in our simulations, the pump propagation is
\begin{equation}
\mathbf{A}(z)=e^{\,-i z H_p}\,\mathbf{A}(0),
\label{eq:pump_propagator}
\end{equation}
where $H_p$ is the pump Hamiltonian.

Let $\{ \mathbf{v}_m \}_{m=1}^N$ be an orthonormal eigenbasis of $H_p$,
\begin{equation}
H_p\,\mathbf{v}_m=\lambda_m\,\mathbf{v}_m,
\qquad
\mathbf{v}_m^\dagger \mathbf{v}_n=\delta_{mn}.
\end{equation}
Assume the pump is launched as a single pump eigenvector,
\begin{equation}
\mathbf{A}(0)=\mathbf{v}_{m_0}.
\end{equation}
Then, using Eq.~\eqref{eq:pump_propagator},
\begin{equation}
\mathbf{A}(z)=e^{\,-izH_p}\mathbf{v}_{m_0}
= e^{\,-iz\lambda_{m_0}}\,\mathbf{v}_{m_0}.
\label{eq:pump_stays_eigenvector}
\end{equation}
Therefore, the pump remains exactly in the same eigenmode for all $z$; only a global phase evolves.

Define the modal overlap used in the numerical code,
\begin{equation}
F_m(z)=\big|\mathbf{v}_m^\dagger \mathbf{A}(z)\big|^2.
\label{eq:F_def}
\end{equation}
Substituting Eq.~\eqref{eq:pump_stays_eigenvector} into Eq.~\eqref{eq:F_def} yields
\begin{equation}
\mathbf{v}_m^\dagger \mathbf{A}(z)
= e^{\,-iz\lambda_{m_0}}\,\mathbf{v}_m^\dagger \mathbf{v}_{m_0}
= e^{\,-iz\lambda_{m_0}}\,\delta_{m m_0},
\end{equation}
and hence the exact identity
\begin{equation}
F_m(z)=\delta_{m m_0}
\label{eq:F_delta}
\end{equation}

\begin{figure}[!h]
    \centering
    \includegraphics[width=1\textwidth]{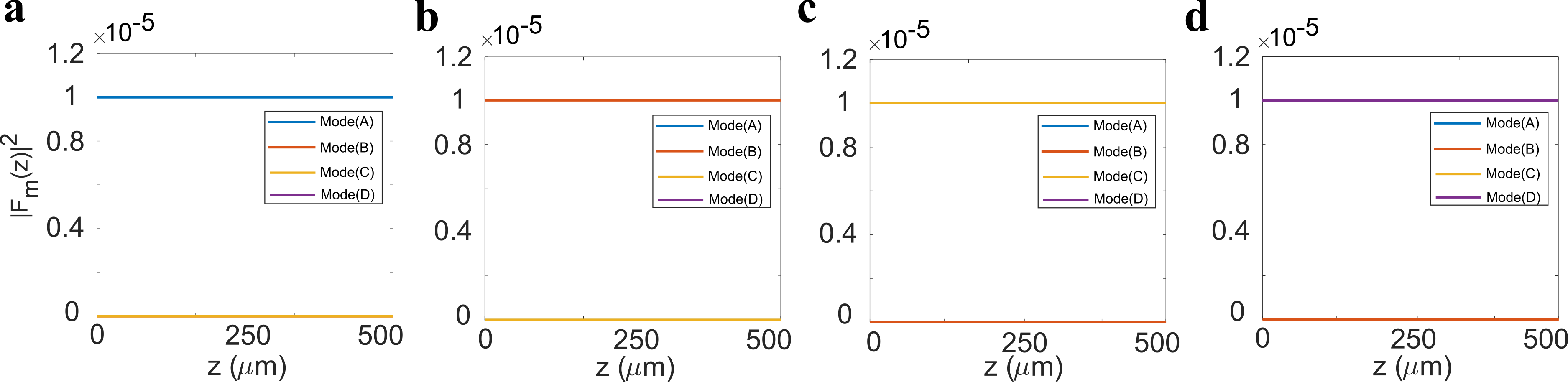}
\caption{\textbf{Pump-eigenmode excitation and conserved modal overlap in a \(J=5\) topological superlattice.}
\textbf{a--d,} Evolution of the normalized pump-eigenmode overlap,
\(|F_m(z)|^2 = |\langle v_m|A(z)\rangle|^2\), along the propagation direction \(z\) for inputs initialized in topological pump modes \(A\)--\(D\), respectively. 
In each case, the launched eigenmode maintains unit overlap throughout propagation, while projections onto the remaining topological modes stay near zero. 
This confirms that exact eigenmode excitation preserves the modal content of the pump field, apart from the accumulated propagation phase.}
    \label{Fig:FJ5}
\end{figure}

Figure~\ref{Fig:FJ5} verifies this eigenmode-selective excitation in
the $J=5$ topological superlattice. When the input pump is prepared in
any one of the four topological eigenmodes (Modes A--D), the
corresponding occupation $F_m(z)$ remains at unity throughout the
device, while the occupations of the remaining topological modes stay
at zero to within numerical roundoff, confirming mixing-free pump
evolution up to an overall accumulated phase.

\vspace{0.5em}
\noindent
\textbf{Eigenmode expansions and projection.}
Diagonalize the signal and idler Hamiltonians,
\begin{equation}
H_sV_s=V_s\Lambda_s,\qquad H_iV_i=V_i\Lambda_i,
\label{eq:S3_diag}
\end{equation}
where $V_s=[s_1,\ldots,s_N]$ and $V_i=[i_1,\ldots,i_N]$ are unitary matrices of orthonormal eigenvectors,
and $\Lambda_s=\mathrm{diag}(\lambda_{s,1},\ldots,\lambda_{s,N})$, $\Lambda_i=\mathrm{diag}(\lambda_{i,1},\ldots,\lambda_{i,N})$.
We define the biphoton amplitude in the joint eigenmode basis by the unitary change of basis
\begin{equation}
\psi_{\mathrm{eig}}(z)=V_s^\dagger\,\psi(z)\,V_i
\qquad\Longleftrightarrow\qquad
\psi_{\mathrm{eig}}(m,n;z)=\langle s_m|\psi(z)|i_n\rangle .
\label{eq:S3_def_eig}
\end{equation}
Substituting $\psi(z)=V_s\psi_{\mathrm{eig}}(z)V_i^\dagger$ into Eq.~\eqref{eq:biphoton_eom} and left-/right-multiplying by
$V_s^\dagger$ and $V_i$ yields
\begin{equation}
i\frac{d}{dz}\psi_{\mathrm{eig}}(z)
=
\Lambda_s\,\psi_{\mathrm{eig}}(z)+\psi_{\mathrm{eig}}(z)\,\Lambda_i
+\gamma\psi_0\,S(z),
\label{eq:S3_matrix_eig}
\end{equation}
where the pump-induced source term in the eigenmode basis is
\begin{equation}
S(z)
=
V_s^\dagger\,\mathrm{diag}\!\big(A_1^2(z),\ldots,A_N^2(z)\big)\,V_i,
\qquad
S_{mn}(z)=\langle s_m|\mathrm{diag}(A^2(z))|i_n\rangle .
\label{eq:S3_source_def}
\end{equation}
Writing $S_{mn}(z)$ explicitly in the site basis gives the overlap form
\begin{equation}
S_{mn}(z)=\sum_{r=1}^{N} s_m^*(r)\,A_r^2(z)\,i_n(r),
\label{eq:S3_source_overlap}
\end{equation}
which shows that a given eigenmode pair $(m,n)$ is driven only to the extent that the pump intensity profile overlaps the spatial product $s_m^*(r)i_n(r)$. Because $\Lambda_s$ and $\Lambda_i$ are diagonal, Eq.~\eqref{eq:S3_matrix_eig} decouples into $N^2$ scalar driven ODEs:
\begin{equation}
i\frac{d}{dz}\psi_{\mathrm{eig}}(m,n;z)
=
\big(\lambda_{s,m}+\lambda_{i,n}\big)\psi_{\mathrm{eig}}(m,n;z)
+\gamma\psi_0\,S_{mn}(z).
\label{eq:S3_scalar}
\end{equation}
Thus, each mode pair behaves like a single driven oscillator with ``two-photon eigenenergy'' $\lambda_{s,m}+\lambda_{i,n}$ and a drive amplitude $S_{mn}(z)$. For the vacuum initial condition $\psi(0)=0$ (equivalently $\psi_{\mathrm{eig}}(0)=0$), Eq.~\eqref{eq:S3_scalar} is solved by variation of constants:
\begin{equation}
\psi_{\mathrm{eig}}(m,n;L)
=
-i\gamma\psi_0
\int_{0}^{L}dz'\,
e^{-i(\lambda_{s,m}+\lambda_{i,n})(L-z')}\,
S_{mn}(z').
\label{eq:S3_integral}
\end{equation}
Equation~\eqref{eq:S3_integral} is exact and shows that the amplitude of the biphotons in channel $(m,n)$ is the accumulated coherent sum of the pump-induced overlap $S_{mn}(z')$, weighted by the phase evolution of the corresponding two-photon eigenenergy.

In the following, we specialize Eq.~\eqref{eq:S3_integral} to the experimentally relevant case where the pump is launched in a single pump eigenmode (or a controlled superposition of two eigenmodes), for which $A(z)$ acquires a simple $z$-dependent phase. This reduces $S_{mn}(z)$ to a known phase factor times a static spatial overlap, and the integral in Eq.~\eqref{eq:S3_integral} can then be analytically evaluated to produce a compact closed-form expression highlighting the phase-mismatch (detuning) and the $L$-dependent sinc-type selectivity.

Now, impose the experimentally relevant condition used in the simulation:
the pump is launched as a pump eigenvector $\mathbf{v}_{m_0}$ (a localized topological pump mode).
From Eq.~\eqref{eq:pump_stays_eigenvector},
\begin{equation}
A_r(z)=e^{-iz\lambda_{m_0}}\,v_{m_0}(r)
\quad\Rightarrow\quad
A_r^2(z)=e^{-i2z\lambda_{m_0}}\,\big[v_{m_0}(r)\big]^2.
\end{equation}
Hence Eq.~\eqref{eq:S3_source_overlap} factorizes as
\begin{equation}
S_{mn}(z)=e^{-i2z\lambda_{m_0}}\,S_{mn}^{(0)},
\qquad
S_{mn}^{(0)}=\sum_{r=1}^N s_m^*(r)\,\big[v_{m_0}(r)\big]^2\,i_n(r).
\label{eq:Smn_factor}
\end{equation}
Substituting Eq.~\eqref{eq:Smn_factor} into Eq.~\eqref{eq:S3_integral} yields a closed form:

\begin{equation}
\psi_{\mathrm{eig}}(m,n;L)
=
\gamma\psi_0\,S_{mn}^{(0)}\,
\frac{1-e^{-i\Delta\beta_{mn}L}}{\Delta\beta_{mn}}\,
e^{-i2\lambda_{m_0}L},
\label{eq:psi_closed_form}
\end{equation}
where the detuning (phase-mismatch) is
\begin{equation}
\Delta\beta_{mn}
=
2\lambda_{m_0}-\lambda_{s,m}-\lambda_{i,n}.
\label{eq:detuning}
\end{equation}
Therefore, the eigenmode-basis population obeys the selectivity law

\begin{equation}
\bigl|\psi_{\mathrm{eig}}(m,n;L)\bigr|^2
\propto
\bigl|S_{mn}^{(0)}\bigr|^2\,
\frac{4\sin^2\!\left(\tfrac{\Delta\beta_{mn}L}{2}\right)}{\left(\Delta\beta_{mn}\right)^2}
\label{eq:selectivity_rule}
\end{equation}

Equation~\eqref{eq:selectivity_rule} shows that biphoton population is controlled by:
(i) the spatial overlap $\big|S_{mn}^{(0)}\big|^2$ between the squared pump eigenvector profile
and the signal/idler eigenmodes; and (ii) phase matching via the detuning $\Delta\beta_{mn}$.

Consequently, when restricting $(m,n)$ to the three localized topological modes $\{A,B,C,D\}$,
the observed 4$\times$4 ``biphoton eigenmode population'' can collapse to a single dominant block (e.g. $DD$)
whenever
\begin{equation}
\big|S_{DD}^{(0)}\big|^2\,\frac{1}{\Delta\beta_{DD}^2}
\gg
\big|S_{mn}^{(0)}\big|^2\,\frac{1}{\Delta\beta_{mn}^2}
\quad\text{for all other }(m,n)\in\{A,B,C,D\}^2,
\end{equation}
and/or when $\Delta\beta_{DD}$ is much closer to zero than the other detunings.
This provides a rigorous analytical explanation for the strongly selective 4$\times$4 population map
observed in the simulation when the pump is launched in topological mode 1.
For short devices or weak phase mismatch ($|\Delta\beta_{mn}|L\ll 1$),
Eq.~\eqref{eq:psi_closed_form} gives
\begin{equation}
\big|\psi_{\mathrm{eig}}(m,n;L)\big|^2
\approx
(\gamma\psi_0)^2\,\big|S_{mn}^{(0)}\big|^2\,L^2,
\end{equation}
so population grows quadratically with length.
For large mismatch ($|\Delta\beta_{mn}|L\gg 1$), the $\sin^2$ factor oscillates and the envelope scales as
$\propto \big|S_{mn}^{(0)}\big|^2/\Delta\beta_{mn}^2$, producing strong selectivity in $(m,n)$.

The biphoton eigenmode operator for a signal/idler eigenmode pair $(m,n)$ is the outer product
\begin{equation}
\hat{B}_{mn} = |\mathbf{s}_m\rangle\langle \mathbf{i}_n|,
\end{equation}
and the corresponding biphoton eigenmode amplitude (overlap) is
\begin{equation}
\mathrm{Overlap}_{mn}
=
\langle \mathbf{s}_m|\psi|\mathbf{i}_n\rangle
=
\sum_{a=1}^N\sum_{b=1}^N s_m^*(a)\,\psi(a,b)\,i_n(b),
\end{equation}
so the eigenmode population is $\lvert \mathrm{Overlap}_{mn}\rvert^2$.
This is exactly the formula implemented numerically via the double-sum projection.

\begin{figure}[!h]
    \centering
    \includegraphics[width=1\textwidth]{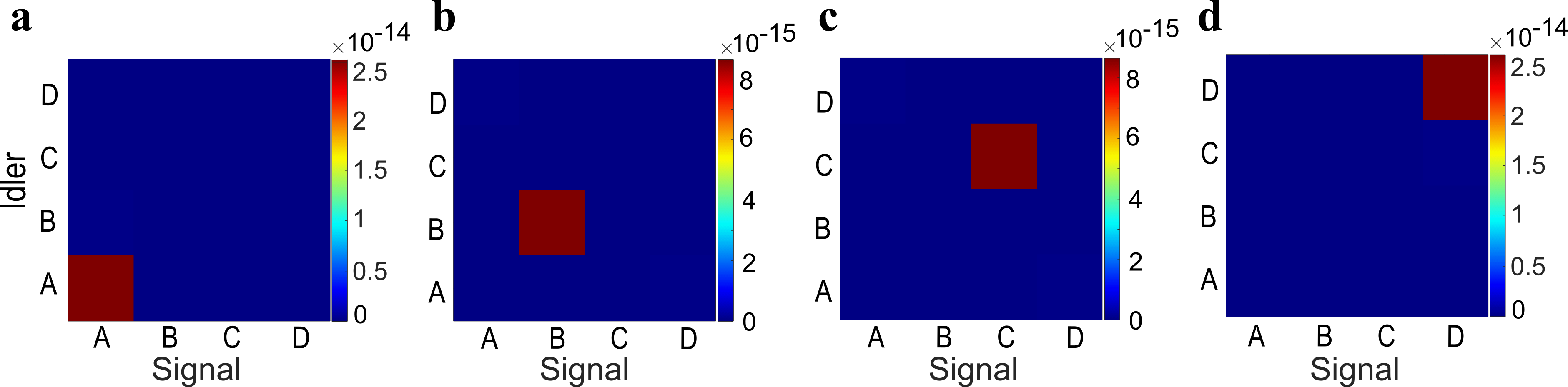}
\caption{\textbf{Biphoton eigenmode populations in a \(J=5\) topological superlattice under eigenmode-selective pumping.}
\textbf{a--d,} Coincidence population matrices, \(P_{mn}=|\langle s_m|\psi(L)|i_n\rangle|^2\), projected onto the four localized topological signal and idler eigenmodes, labelled \(A\)--\(D\). In each panel, the pump is launched into the corresponding topological pump eigenmode. The generated population is concentrated predominantly in the matched mode-pair block, demonstrating selective addressing of individual channels within the \(J=5\) topological biphoton subspace.}
    \label{Fig:J5}
\end{figure}

Figure~\ref{Fig:J5} shows the biphoton state projected onto the topological eigenmode subspace (A--D) for a $J=5$ device when the pump is launched in different topological pump eigenvectors. In each case, the biphoton eigenmode population matrix is strongly concentrated in the block associated with the launched mode, while the remaining blocks stay near the noise floor, indicating that the nonlinear generation predominantly couples into the corresponding topological signal--idler mode pair and that cross-population into other topological channels is strongly suppressed.

\suppheading{S3. Selective quantum-state generation with topological pump-mode excitation}

\begin{figure}[H]
    \centering
    \includegraphics[width=1\textwidth]{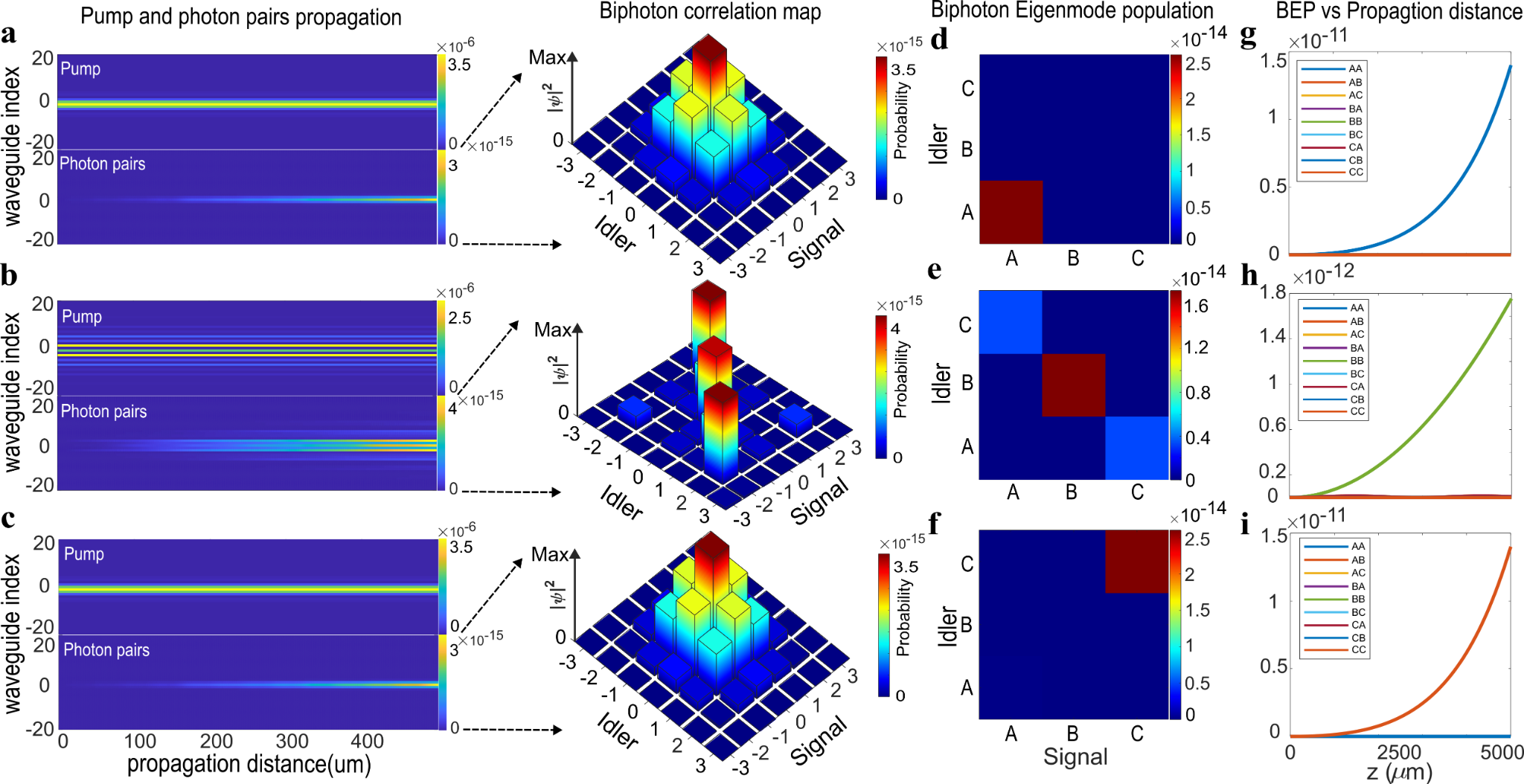}
\caption{\textbf{Simulations of the nonlinear generation of quantum states in a \(J=4\) superlattice under selective excitation of individual topological modes.}
\textbf{a--c,} Excitation of the three localized topological pump eigenmodes and the corresponding pump evolution, photon-pair generation and output biphoton correlation maps within the seven-waveguide interface region. 
\textbf{d--f,} Biphoton eigenmode population matrices obtained by projecting the generated states onto the localized topological signal--idler eigenmode basis. Each pump eigenmode preferentially populates a distinct biphoton eigenchannel, establishing a direct mapping between the pump topology and the generated quantum state.
\textbf{g--i,} Propagation-dependent evolution of the biphoton eigenmode populations. In each case, the selected eigenchannel dominates the nonlinear generation process, whereas non-target channels remain strongly suppressed. These results demonstrate programmable, high-selectivity generation of localized topological biphoton states.}
    \label{J=4}
\end{figure}

\begin{figure}[H]
    \centering
    \includegraphics[width=1\textwidth]{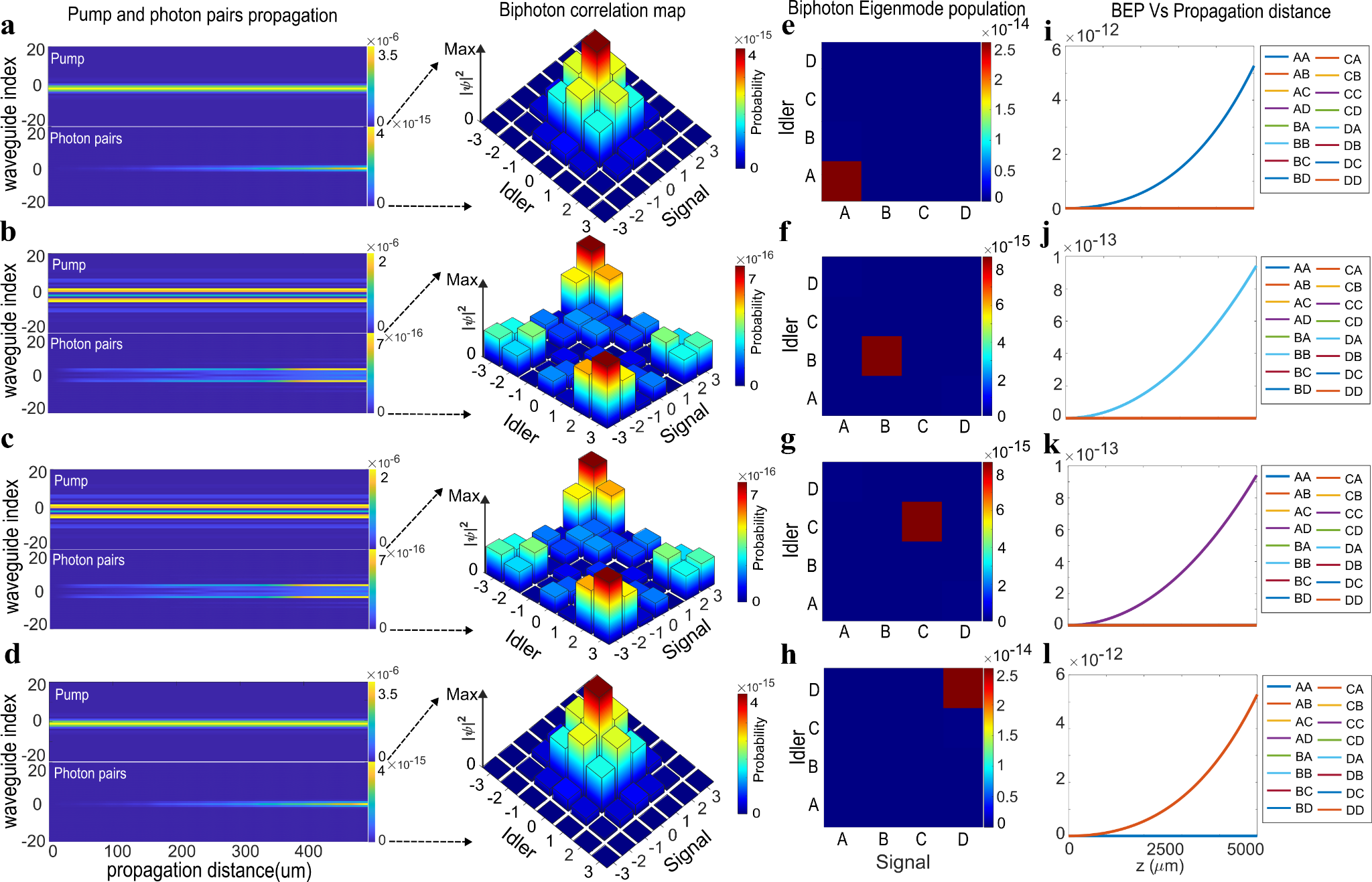}
\caption{\textbf{Simulations of the nonlinear generation of quantum states in a \(J=5\) superlattice under selective excitation of individual topological modes..}
\textbf{a--d,} Excitation of the four localized topological pump eigenmodes and the corresponding pump evolution, photon-pair generation and output biphoton correlation maps within the seven-waveguide interface region. 
\textbf{e--h,} Biphoton eigenmode population matrices obtained by projecting the generated states onto the localized topological signal and idler eigenmode basis. Each pump eigenmode selectively populates a distinct biphoton eigenchannel, enabling deterministic state selection. 
\textbf{i--l,} Propagation-dependent evolution of the biphoton eigenmode populations, showing preferential growth of the targeted channel while competing channels remain suppressed. The paired responses of modes 1 and 4 and of modes 2 and 3 reflect the spectral symmetry of the \(J=5\) topological superlattice.}
    \label{J=5}
\end{figure}

\begin{figure}[H]
    \centering
    \includegraphics[width=1\textwidth]{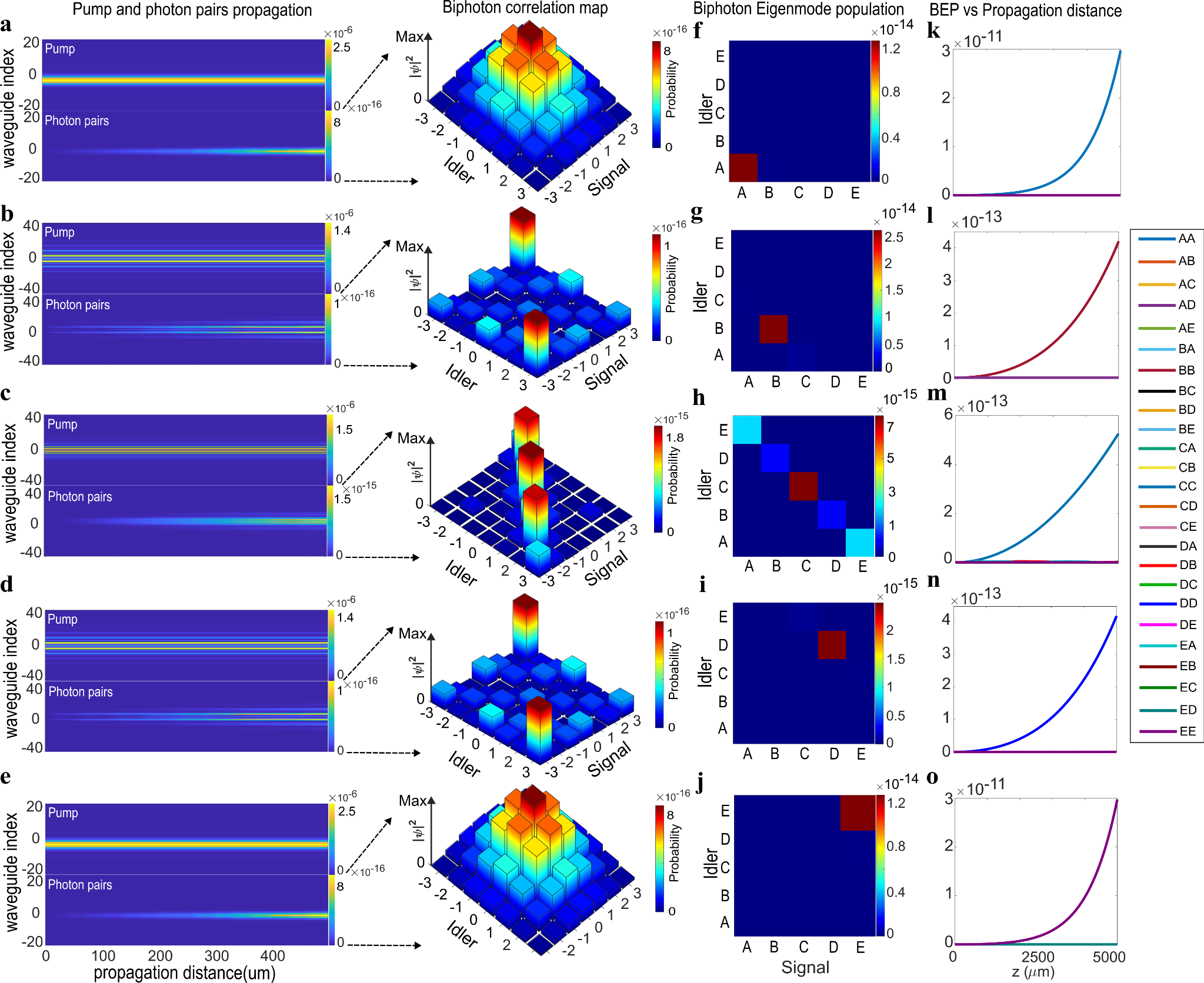}
\caption{\textbf{Simulations of the nonlinear generation of quantum states in a \(J=6\) superlattice under selective excitation of individual topological modes.}
\textbf{a--e,} Excitation of the five localized topological pump eigenmodes and the corresponding pump evolution, photon-pair generation and output biphoton correlation maps within the seven-waveguide interface region. 
\textbf{f--j,} Biphoton eigenmode population matrices obtained by projecting the generated states onto the localized topological signal and idler eigenmode basis. Excitation of each pump eigenmode selectively populates a distinct biphoton eigenchannel, demonstrating deterministic control over the generated entangled state. 
\textbf{k--o,} Propagation-dependent evolution of the biphoton eigenmode populations. In each case, the targeted eigenchannel is preferentially amplified during propagation, whereas non-target channels remain weak. 
Together, these results show that increasing the superlattice order enlarges the number of accessible topological quantum channels while preserving strong state selectivity, providing a scalable route to high-dimensional topological entanglement engineering.}
    \label{J=6}
\end{figure}

We next investigate selective quantum-state generation by exciting individual localized topological pump modes of the superlattice. In Figs.~\ref{J=4}--\ref{J=6}, the pump is launched as the full topological eigenvector of the corresponding pump Hamiltonian. This ideal eigenmode excitation provides a reference for the intrinsic modal selectivity of the system, independent of finite-aperture experimental constraints. Because the topological modes are spatially localized at the interface, each launched pump mode preferentially drives four-wave-mixing generation within a well-defined localized signal--idler eigenmode channel. As a result, the generated biphoton state can be programmed by choosing which topological pump eigenmode is excited.

Figures~\ref{J=4}, \ref{J=5}, and \ref{J=6} show the selective generation dynamics for the $J=4$, $J=5$, and $J=6$ superlattices, respectively. Each column corresponds to one selected topological pump mode. For each excitation, the upper panel shows the pump intensity evolution along the array, the middle panel shows the output biphoton correlation map restricted to the seven central waveguides, and the lower panel shows the photon-pair intensity evolution during propagation. In all cases, the generated photon-pair correlations remain strongly concentrated near the interface, confirming that the nonlinear interaction is confined to the localized topological region. Importantly, changing the launched pump eigenmode produces a systematic redistribution of the biphoton correlation pattern, demonstrating mode-dependent quantum-state generation.

The modal selectivity is quantified by projecting the generated two-photon state onto the localized topological signal--idler eigenmode basis. The biphoton eigenmode population is defined as
\[
P_{mn}(z)=\left|\left\langle v^{(s)}_m v^{(i)\,*}_n \middle| \psi(z) \right\rangle\right|^2 ,
\]
where $v^{(s)}_m$ and $v^{(i)}_n$ are the localized topological signal and idler eigenmodes, respectively, and $\psi(z)$ is the generated biphoton wavefunction. The eigenmode population matrices show that the dominant population appears in the signal--idler channel associated with the selected pump mode, while off-target modal channels remain strongly suppressed. The corresponding propagation-dependent biphoton eigenmode populations further confirm that the targeted biphoton channel grows selectively along the device.

For the $J=5$ superlattice, the responses associated with modes 1 and 4 are nearly identical, as are those associated with modes 2 and 3. This pairing is a consequence of the symmetry of the localized topological spectrum, which organizes the interface states into symmetry-related modal partners. Therefore, the selective excitation scheme does not merely address spatially localized states; it also accesses symmetry-structured quantum channels within the topological biphoton subspace.

\begin{figure}[H]
    \centering
    \includegraphics[width=\textwidth]{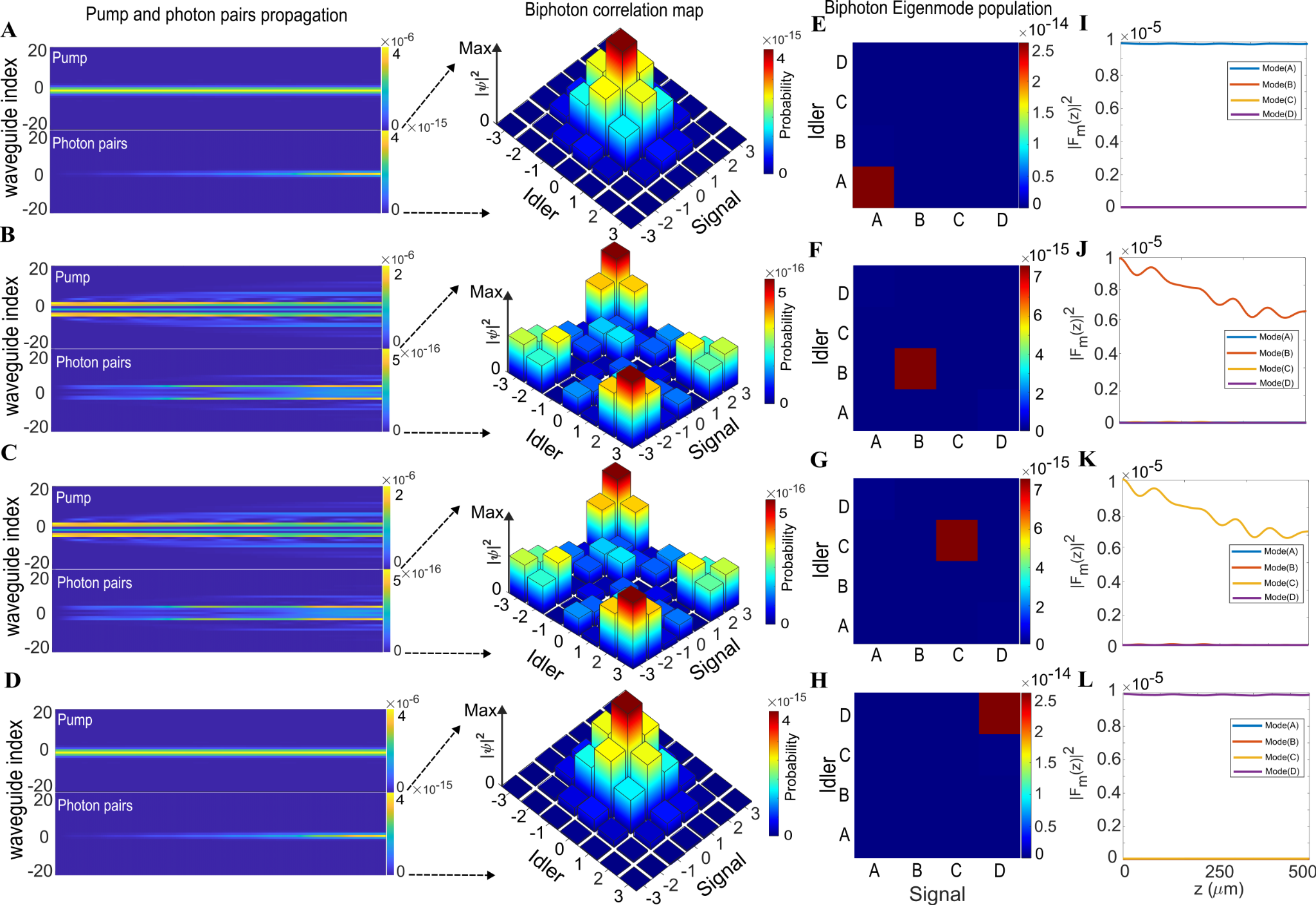}
\caption{\textbf{Simulations of the nonlinear generation of quantum states in a \(J=5\) superlattice under truncated pump excitation.}
Each pump input is constructed from the five central interface sites of the corresponding topological pump eigenmode, rather than from the full ideal eigenvector used in Fig.~\ref{J=5}. This truncation emulates the finite spatial aperture of the multi-port excitation scheme implemented experimentally.
\textbf{a--d,} Propagation dynamics for truncated excitation of topological pump modes 1--4. The panels show the pump evolution, photon-pair generation and output biphoton correlation distributions within the seven-waveguide interface region, \(-3 \leq n \leq 3\), with \(n=0\) denoting the interface site.
\textbf{e--h,} Biphoton eigenmode population matrices,
\(P_{mn}=\left|\left\langle v^{(s)}_m \left(v^{(i)}_n\right)^* \middle| \psi(L)\right\rangle\right|^2\),
obtained by projecting the generated state onto the localized topological signal--idler eigenmode basis. Despite the strong truncation of the pump profile, the generated states remain dominated by the intended biphoton eigenchannels, demonstrating robust eigenmode selectivity.
\textbf{i--l,} Evolution of the pump-eigenmode projections,
\(\left|\langle v^{(p)}_m | A(z) \rangle\right|^2\),
during propagation. The truncated inputs retain strong overlap with their corresponding ideal topological pump eigenmodes, explaining the preservation of selective biphoton generation. The agreement with the full-eigenvector results of Fig.~\ref{J=5} shows that excitation through only five interface waveguides is sufficient to reproduce the desired topological quantum-state-engineering protocol.}
    \label{semi5}
\end{figure}

Finally, Fig.~\ref{semi5} connects the ideal full-eigenvector excitation scheme to an experimentally realistic finite-aperture pump configuration. In this figure only, the pump is not launched as the full eigenvector of the entire lattice. Instead, each $J=5$ topological pump mode is approximated by retaining only its five dominant interface components. Here, the notation $K=5$ refers to the five-site truncation aperture of the pump profile, not to the Schmidt number. This experimentally motivated truncation reflects the practical situation in which only the central input waveguide and its two nearest neighbors on each side are directly addressed. The strong agreement between Fig.~\ref{J=5} and Fig.~\ref{semi5} shows that the essential selective-generation physics is preserved under finite-aperture excitation: the pump remains concentrated in the targeted topological channel, the biphoton correlations remain localized near the interface, and the dominant signal--idler eigenmode population follows the selected pump mode. This comparison confirms that experimentally accessible truncated pump profiles are sufficient to reproduce the ideal eigenmode-selective quantum-state generation enabled by localized topological modes.

\suppheading{S4. Experimental implementation of the excitation and the quantum measurements}

\begin{figure}[H]
    \centering
    \includegraphics[width=0.85\textwidth]{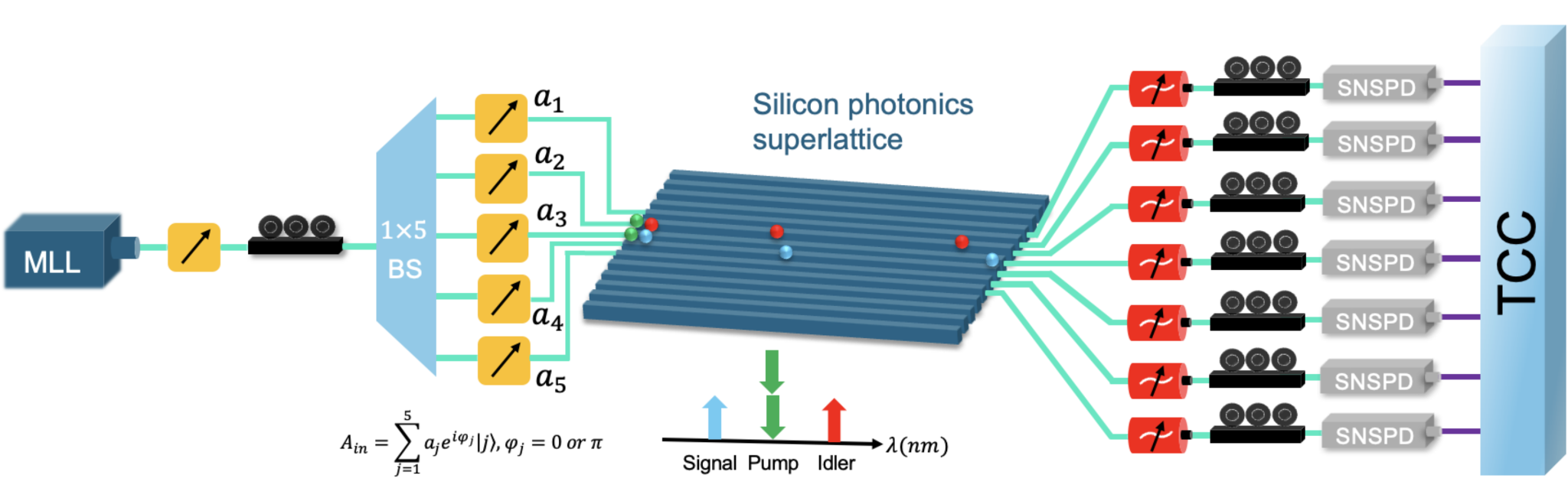}
    \caption{
 \textbf{Experimental scheme for programmable excitation and site-resolved detection of topological biphoton states in a silicon photonics superlattice. }
A sub-picosecond pump at $1550~\mathrm{nm}$ is power- and polarization-controlled, coherently divided into five input channels, and independently weighted by per-channel tunable attenuators to define the amplitudes $a_{1-5}$. 
The five programmed inputs address the central interface waveguide and the two nearest waveguides on each side, while integrated on-chip phase shifters impose the relative phases $\varphi_j=0$ or $\pi$, preparing $A_{\mathrm{in}}=\sum_{j=1}^{5}a_j e^{i\varphi_j}|j\rangle$. 
The tailored pump excites topological interface modes and drives spontaneous four-wave mixing, generating spatially structured signal--idler photon pairs. 
The seven central output waveguides are collected, spectrally filtered to separate signal photons near $1545~\mathrm{nm}$ and idler photons near $1555~\mathrm{nm}$ while rejecting the residual pump, and detected with SNSPDs. 
Coincidence events recorded by the time-correlation circuit reconstruct the biphoton correlation matrix.
}
    \label{exp-seup}
\end{figure}

Sub-picosecond pulses at $1550~\mathrm{nm}$ from a mode-locked laser with 40 MHz repetition rate are power split into five fibers and coupled via an optical fiber array into the five waveguides located at the center of the topological interface  using grating coupler, as illustrated in Fig.~\ref{exp-seup}. The initial pump power and polarization was optimized to excite the desired on-chip mode by adding tunable attenuator and polarization controller after laser. Each channel contains a tunable attenuator and an on-chip phase shifter, enabling simultaneous control of the amplitude and phase of the launched field. The attenuators set the relative pump weights, while the phase shifters implement the required sign of each coefficient through either a zero or $\pi$ phase shift. The prepared input field can therefore be written as $A_{\mathrm{in}}=\sum_{j=1}^{5}a_j e^{i\varphi_j}|j\rangle$, where $a_j$ is the programmed input amplitude, $\varphi_j=0$ or $\pi$ defines the relative sign, and $|j\rangle$ denotes the excited input waveguide. This architecture provides a direct experimental route to synthesize the real-valued topological modal superpositions used in the optimization. The five-channel architecture therefore provides the full control space required for the largest modal manifold considered here, while lower-dimensional cases are obtained by addressing the corresponding subset of channels. 

As the pulses propagate through the structure, their high peak power drives spontaneous four-wave mixing (SFWM), generating energy-time entangled photon pairs that inherit the spatial and symmetry structure of the underlying topological modes. At the end of the lattice, the seven central waveguides are fanned out and coupled into seven separate single-mode fibers, enabling parallel, site-resolved measurements of the emitted photon pairs. The output photons are spectrally separated using a wavelength division multiplexer with output channeles centered at $1545~\mathrm{nm}$ for the signal photons and approximately $1555~\mathrm{nm}$ for the idler photons. Three additional filetring stages not shown in Fig.~\ref{exp-seup} $-$ 2 before the chip and one after $-$ suppressing residual pump leakage by ~120 dB. The filtered photons are detected using superconducting nanowire single-photon detectors (SNSPDs), and coincidence events are registered with a high-resolution time-correlation circuit (TCC). This measurement procedure reconstructs the spatial probability distribution of the biphoton wavefunction through pairwise correlations between output sites, allowing direct comparison of the measured correlation structure across nominally identical devices and under controlled variations of the excitation conditions.

We extracted the coincidences to accidentals ratio (CAR) from the time-correlation histograms given by the TCC using equal-width integration windows centered on the true coincidence peak and on the adjacent accidental peak, separated by the laser repetition period. The integrated counts in these windows are denoted \(C\) and \(A\), respectively, and the CAR was calculated as \(\mathrm{CAR}=C/A\). Assuming independent Poisson counting statistics, the corresponding uncertainties are \(\sigma_C=\sqrt{C}\) and \(\sigma_A=\sqrt{A}\), with the CAR uncertainty obtained by standard error propagation. A channel was considered populated only when its net coincidence signal exceeded the accidental background by at least \(3\sigma\), corresponding to \((C-A)/\sqrt{C+A}\geq3\). 

\suppheading{S5. Eigenmode-resolved SFWM in a topological superlattice: closed-form pair-channel selectivity and propagation scaling.}

In our lattice SFWM model, the biphoton wavefunction $\psi(z)$ is generated by an on-site nonlinear source proportional to the squared pump field profile, while the signal and idler propagate under their respective tight-binding Hamiltonians. This structure makes the eigenmode basis the natural representation: after diagonalizing the signal and idler Hamiltonians, each signal--idler eigenmode pair defines an independently driven generation channel. As a result, the emergence of a dominant block (e.g., AA, BB, or CC) under eigenmode-selective pumping follows from a compact overlap-and-detuning principle rather than from numerical coincidence.

Diagonalizing $H_s$ and $H_i$ and expanding $\psi(z)$ in the corresponding eigenmode product basis yields a driven first-order equation for each pair amplitude $\psi_{\mathrm{eig}}(m,n;z)=\langle s_m|\psi(z)|i_n\rangle$, with a pump-induced drive
$S_{mn}(z)=\langle s_m|\mathrm{diag}(A_p^2(z))|i_n\rangle$
(\eqref{eq:S3_source_def}, \eqref{eq:S3_scalar}, \eqref{eq:S3_integral}). With the initial condition $\psi(0)=0$, the solution is an exact convolution integral between the free propagation phase of the $(m,n)$ channel and the drive (Eq.~(13)). When the pump is launched as an exact pump eigenvector,
$A_p(z)\propto e^{i\lambda_{p,\alpha}z}v_{p,\alpha}$,
the drive acquires a single longitudinal phase factor and its spatial content reduces to a constant channel-coupling coefficient
\begin{equation}
\kappa_{mn}^{(\alpha)}=\sum_{r=1}^{N} s_m^*(r)\,v_{p,\alpha}^2(r)\,i_n(r),
\end{equation}
so that each eigenmode-pair channel is governed by (i) the mode-selective spatial overlap $|\kappa_{mn}^{(\alpha)}|^2$ and (ii) an effective phase-mismatch
\begin{equation}
\Delta_{mn}^{(\alpha)}=2\lambda_{p,\alpha}-\big(\lambda_{s,m}+\lambda_{i,n}\big),
\end{equation}
which controls the longitudinal accumulation (cf. Eq.~(13)). In particular, the closed-form solution implies a population scaling of the form
\begin{equation}
P_{mn}(L)=|\psi_{\mathrm{eig}}(m,n;L)|^2 \propto |\kappa_{mn}^{(\alpha)}|^2\,
\frac{\sin^2\!\big(\Delta_{mn}^{(\alpha)}L/2\big)}{\big(\Delta_{mn}^{(\alpha)}\big)^2}.
\end{equation}
The near-resonant ($|\Delta_{mn}^{(\alpha)}|L\ll 1$) behavior follows immediately from the small-argument expansion $\sin x \approx x$:
\begin{equation}
\frac{\sin^2\!\big(\Delta L/2\big)}{\Delta^2}
\;\approx\;
\frac{\big(\Delta L/2\big)^2}{\Delta^2}
\;=\;\frac{L^2}{4},
\qquad (|\Delta|L\ll 1),
\end{equation}
so that the dominant, quasi-phase-matched channel grows quadratically with propagation length,
\begin{equation}
P_{mn}(L)\;\propto\;|\kappa_{mn}^{(\alpha)}|^2\,L^2
\qquad (|\Delta_{mn}^{(\alpha)}|L\ll 1),
\end{equation}
producing the observed near-monotonic increase over the device length. In contrast, larger detuning suppresses the accumulated population through the $\big(\Delta_{mn}^{(\alpha)}\big)^{-2}$ factor and can introduce oscillatory accumulation via the $\sin^2$ term. The overall scale is set by the nonlinear prefactor $|\gamma\psi_0|^2$ together with the pump normalization (since $v_{p,\alpha}$ is scaled by the chosen input power). Consequently, eigenmode pumping produces a sparse eigenmode-population map: the block(s) whose signal and idler modes are most strongly supported on the pumped topological region maximize $|\kappa_{mn}^{(\alpha)}|^2$, while cross-block channels are strongly suppressed by small spatial overlap (and typically larger detuning).

\begin{figure}[H]
    \centering
    \includegraphics[width=1\textwidth]{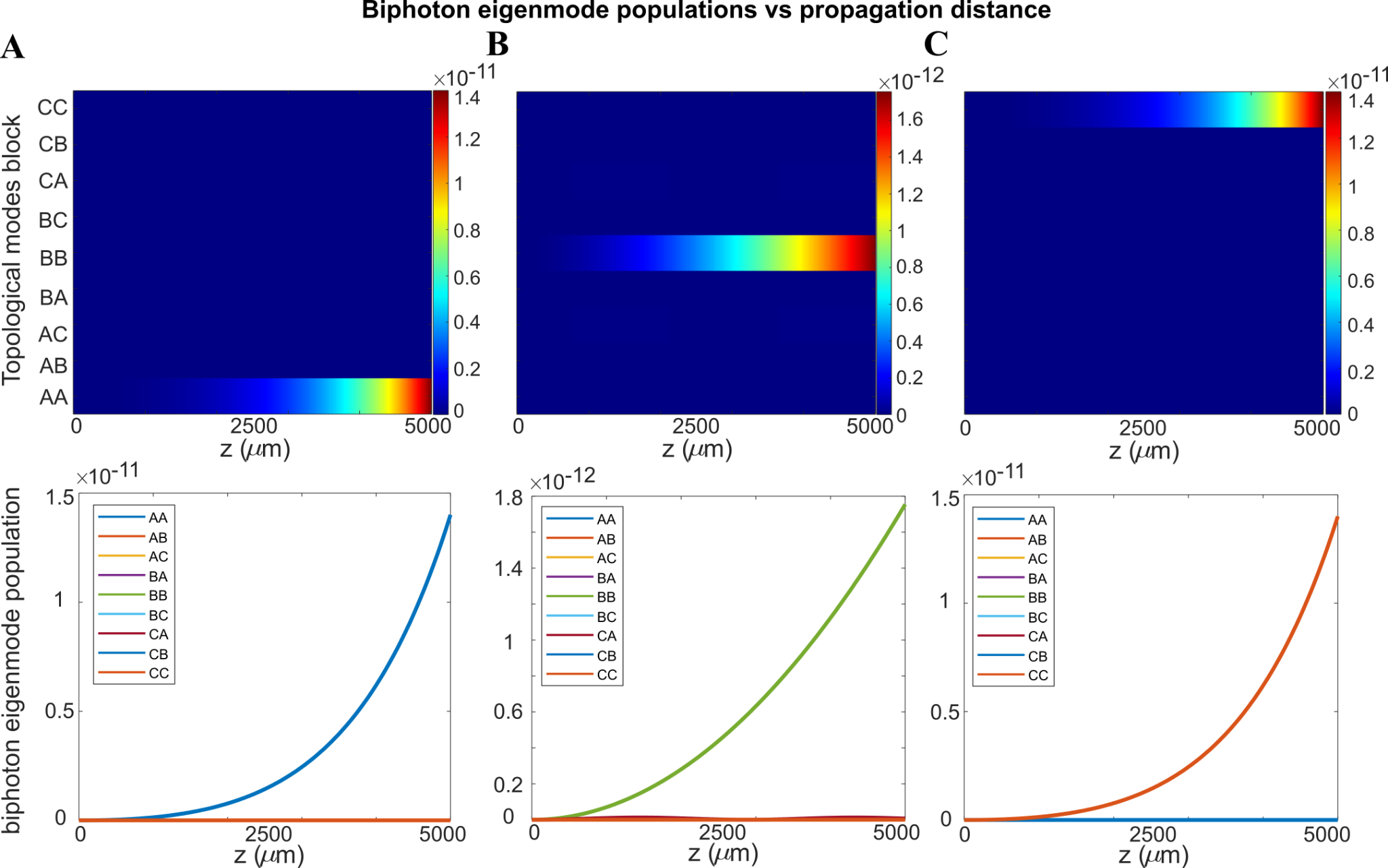}
\caption{\textbf{Biphoton eigenmode populations in a \(J=4\) topological superlattice.}
\textbf{a--c,} Evolution of the nine signal--idler topological mode-pair populations, \(AA\)--\(CC\), as a function of propagation distance \(z\) up to \(5\,\mathrm{mm}\). In each panel, the pump is initialized in a single topological eigenmode, \(A(0)\propto v^{(p)}_{\alpha}\), with \(\alpha\in\{A,B,C\}\). The upper row shows heat maps of the mode-pair populations versus \(z\), while the lower row shows the corresponding line traces for all nine channels. In each case, one biphoton eigenchannel grows monotonically and dominates the generated state, whereas the remaining mode-pair channels stay strongly suppressed, demonstrating high-selectivity topological biphoton generation.}

    \label{Fig:propagation}
\end{figure}

As shown in Fig.~\ref{Fig:propagation}, initializing the pump in a single topological eigenmode produces a highly selective biphoton response in which the population is funneled into one dominant signal--idler product channel while the other eight blocks remain strongly suppressed across the full propagation length. In particular, the dominant block switches with the launched pump eigenvector (here yielding predominantly CC in (A), BB in (B), and AA in (C)), consistent with the fact that the effective generation strength for each channel is set by the spatial overlap coefficient $\kappa_{mn}^{(\alpha)}$ together with the corresponding detuning $\Delta_{mn}^{(\alpha)}$. The observed near-monotonic build-up of the selected block over millimeter-scale propagation is therefore the direct manifestation of quasi-phase-matched accumulation (the small-$|\Delta|L$ regime where the population grows approximately as $L^{2}$), whereas the residual blocks stay near zero because they are simultaneously penalized by weak overlap and/or larger detuning.

When the pump is launched as a generic pulsed spatial profile rather than an exact pump eigenvector, the propagation no longer preserves a single modal component: the field decomposes into multiple pump eigenmodes and continuously beats in the lattice. Figure~\ref{Fig:propagation} therefore probes a physically relevant “non-ideal” regime in which SFWM is driven by a pump that explores the topological region through its evolving intensity pattern, instead of remaining locked to one topological pump eigenstate.

The long-distance dominance of the diagonal blocks (AA, BB, CC) follows from the eigenmode-basis source term $S_{mn}(z)=\langle s_m|\mathrm{diag}(A_p^2(z))|i_n\rangle$, which weights pair generation by the local pump intensity squared and the spatial co-localization of the signal and idler eigenmodes. For topological triplets, modes with the same label on the two arms (A with A, B with B, C with C) are localized on the same physical region and therefore maximize the overlap integral $\kappa_{mm}(z)\sim\sum_r s_m^*(r)\,A_r^2(z)\,i_m(r)$, while cross blocks (e.g., AB, AC, $\ldots$) are suppressed by reduced spatial overlap due to localization at different sites/sublattices. In addition, the accumulated population in each channel scales as
$P_{mn}(L)\propto |\kappa_{mn}|^2\,\sin^2(\Delta_{mn}L/2)/\Delta_{mn}^2$,
so the channels closest to quasi-phase-matching (small $|\Delta_{mn}|$) build up approximately as $L^2$ over the device length, whereas more detuned channels saturate or remain small. Consequently, even though the non-eigenmode pump simultaneously drives several channels, the combination of co-localization and small-detuning bias funnels the biphoton weight predominantly into AA, BB, and CC at large $z$, with the remaining off-diagonal blocks staying near the background level.

\suppheading{S6. Robustness of the selective topological modes entanglement.}

\begin{figure}[H]
    \centering
    \includegraphics[width=1\textwidth]{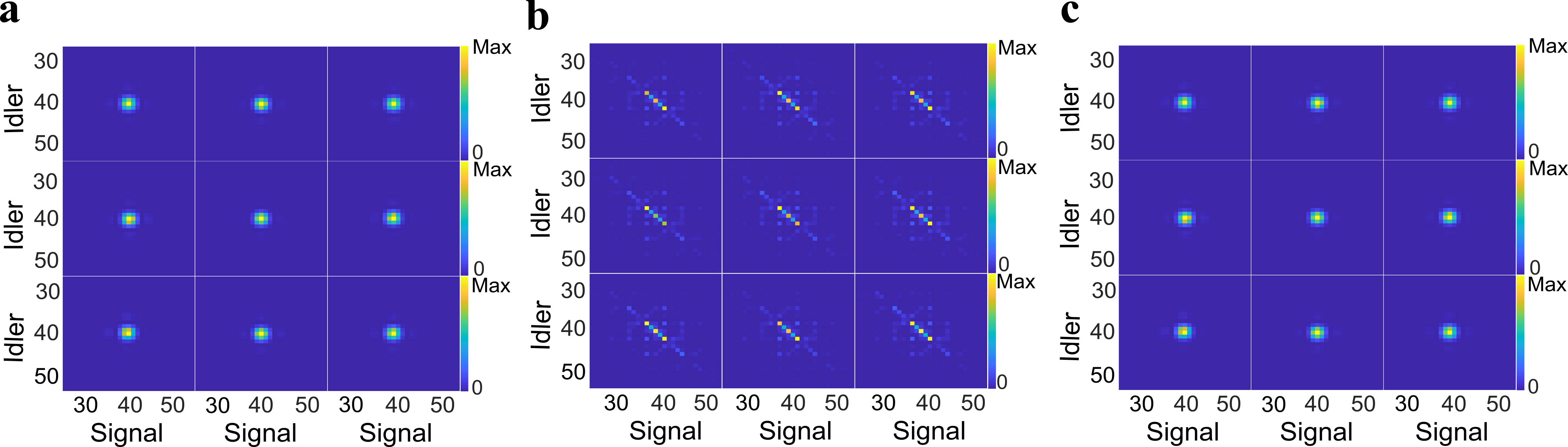}
\caption{\textbf{Biphoton correlations in a disordered \(J=4\) topological superlattice.}
Simulated output biphoton intensity correlations, \(P_{mn}=|\Psi_{mn}|^{2}\), for nine independent realizations of Gaussian coupling disorder with \(\sigma=10\%\) of the nominal coupling strength. Each coincidence map is shown as a function of the signal and idler site indices and is normalized to its own maximum. 
\textbf{a--c,} Results for input excitation of topological pump modes 1, 2 and 3, respectively. The persistence of the dominant correlation features across disorder realizations highlights the robustness of selectively excited localized topological modes.}

    \label{Fig:biphoton}
\end{figure}

When the nonlinear interaction is driven through a specific topological eigenmode, the biphoton state inherits the spectral isolation and real-space localization of that mode, providing a direct mechanism for disorder-resilient quantum correlations. Figure~\ref{Fig:biphoton} illustrates this mode-selective robustness by preparing the pump field as the eigenvector of the topological mode \(1\), \(2\) or \(3\) of supperlattice J=4 and evaluating the output biphoton intensity correlations \(P_{mn}=|\Psi_{mn}|^{2}\) in nine independent realizations of the coupling disorder drawn from a zero-mean Gaussian distribution with \(\sigma=10\%\) of the nominal couplings. Across all disorder realizations, the coincidence weight remains strongly confined to the targeted topological modes and is dominated by the same sparse set of bright correlation peaks associated with the addressed mode, indicating that the pair-generation pathways remain effectively locked to the selected eigenmodes. This behavior is consistent with the fact that random coupling perturbations couple most efficiently between nearly degenerate extended states, whereas the selectively pumped topological modes are both exponentially localized and separated by a gap from the bulk manifold; consequently, leakage into bulk channels is suppressed to higher-order mixing and does not qualitatively alter the correlation structure.

Fabrication tolerances translate into random fluctuations of the
nearest-neighbor coupling rates, which constitute the dominant
chiral-symmetry-preserving perturbation to the engineered coupling
landscape. We characterize their effect on the generated biphoton
state through two complementary, basis-independent metrics: a state
fidelity that quantifies the stability of the output relative to the
disorder-free reference, and the Schmidt number $K$, which measures
the effective entanglement dimensionality. The two are deliberately
chosen to be sensitive to different failure modes---fidelity to any
deformation of the output state, $K$ to the redistribution of
amplitude into additional mode pairs---so that their joint behavior
discriminates between a coherent reshaping of the dominant channel and
a genuine fragmentation of the entanglement structure.

We model the disorder by perturbing each coupling independently,
$t_j \rightarrow t_j(1+\sigma\,\xi_j)$ with $\xi_j$ zero-mean
unit-variance Gaussian variates and $\sigma$ the rms disorder
strength, and average all quantities over an ensemble of
$N_{\mathrm{r}}$ realizations. The disorder-averaged fidelity is the
squared overlap between the unperturbed biphoton wavefunction $\Psi_0$
and its disordered counterpart $\Psi_\sigma$,
\begin{equation}
\mathcal{F}(\sigma)=\Big\langle
\big|\langle \Psi_0 | \Psi_\sigma\rangle\big|^{2}
\Big\rangle_{\!\mathrm{dis}},
\qquad
\langle \Psi|\Phi\rangle=\sum_{m,n}\Psi^{*}_{mn}\Phi_{mn},
\label{eq:fidelity_overlap}
\end{equation}
with both states normalized in the two-photon site basis
$\{|m,n\rangle\}$. In parallel, we compute the Schmidt number $K$ from
the singular values of the biphoton amplitude matrix $\Psi_\sigma$, a
basis-independent measure of how many orthogonal signal--idler mode
pairs contribute appreciably to the state.

\begin{figure}[H]
    \centering
    \includegraphics[width=1\textwidth]{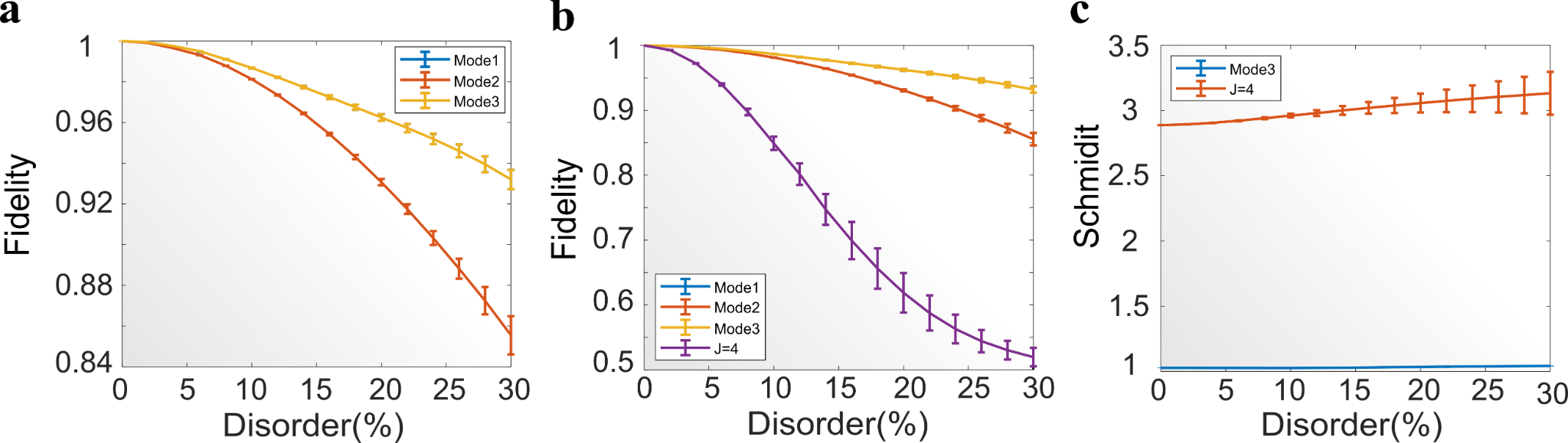}
\caption{\textbf{Disorder dependence of biphoton correlations in the \(J=4\) superlattice.}
\textbf{a,} Biphoton-state fidelity as a function of coupling disorder for selective excitation of topological modes 1--3. Modes 1 and 3 exhibit nearly identical, higher robustness because of their similar localization profiles, whereas mode 2 degrades more rapidly owing to its stronger concentration near the short-gap region.
\textbf{b,} Fidelity comparison between mode-selective pumping and non-selective excitation of the \(J=4\) superlattice. Addressing an isolated topological eigenchannel suppresses disorder-induced coupling to unwanted pathways and preserves the output correlation pattern over a broader disorder range.
\textbf{c,} Schmidt number as a function of disorder for selective excitation of mode 3 and for non-selective \(J=4\) excitation. Mode-selective pumping remains close to the single-channel limit, \(K\simeq1\), whereas non-selective excitation accesses the three-mode topological manifold.}

    \label{Fig:fidelity}
\end{figure}

Figure~\ref{Fig:fidelity} summarizes the disorder response of the
$J=4$ superlattice. When the pump is prepared as a single topological
eigenvector [Fig.~\ref{Fig:fidelity}(A)], $\mathcal{F}(\sigma)$ decays
only gradually, confirming that the output remains locked to the
addressed eigenchannel well beyond realistic fabrication tolerances.
Modes~1 and~3 exhibit nearly identical robustness, consistent with
their comparable localization length: disorder couples both to the
remaining spectrum through similarly weak matrix elements, so their
overlap with the disorder-free target erodes at the same rate.
Mode~2 degrades more rapidly because its amplitude is concentrated in
the short-gap region, where the larger coupling scale converts a fixed
relative disorder $\sigma$ into larger absolute coupling fluctuations;
the resulting hybridization with neighboring states produces a
correspondingly stronger deformation of the biphoton wavefunction.

The benefit of selective eigenmode addressing is made explicit in
Fig.~\ref{Fig:fidelity}(B), which contrasts single-mode operation with
a non-selective excitation that drives several topological modes
simultaneously. The non-selective state degrades markedly faster:
populating multiple spectral components opens many near-resonant
channels through which disorder redistributes amplitude via
bulk-mediated mixing. Eigenmode addressing instead confines the
nonlinear drive to a spectrally isolated, spatially localized channel,
parametrically suppressing disorder-induced leakage into the extended
bulk states and preserving the wavefunction overlap to substantially
higher $\sigma$.

Figure~\ref{Fig:fidelity}c connects this robustness to the modal
content of the state. Under mode-selective pumping (shown for mode~3),
$K$ remains pinned near unity across the full disorder range, so the
biphoton state stays effectively single-channel rather than
fragmenting into independent mode pairs. The non-selective excitation
yields a larger $K$---consistent with the simultaneous participation
of three topological modes---that grows with $\sigma$ as additional
pathways outside the intended topological subspace are progressively
activated. Taken together, the fidelity and the Schmidt number
establish that selective excitation preserves not only the spatial
correlation pattern but also the dimensionality of the underlying
entanglement, the quantity of direct relevance to its use as a
quantum resource.

\suppheading{S7. Finite-aperture demultiplexing of topological modes}

\begin{figure}[H]
    \centering
    \includegraphics[width=0.7\textwidth]{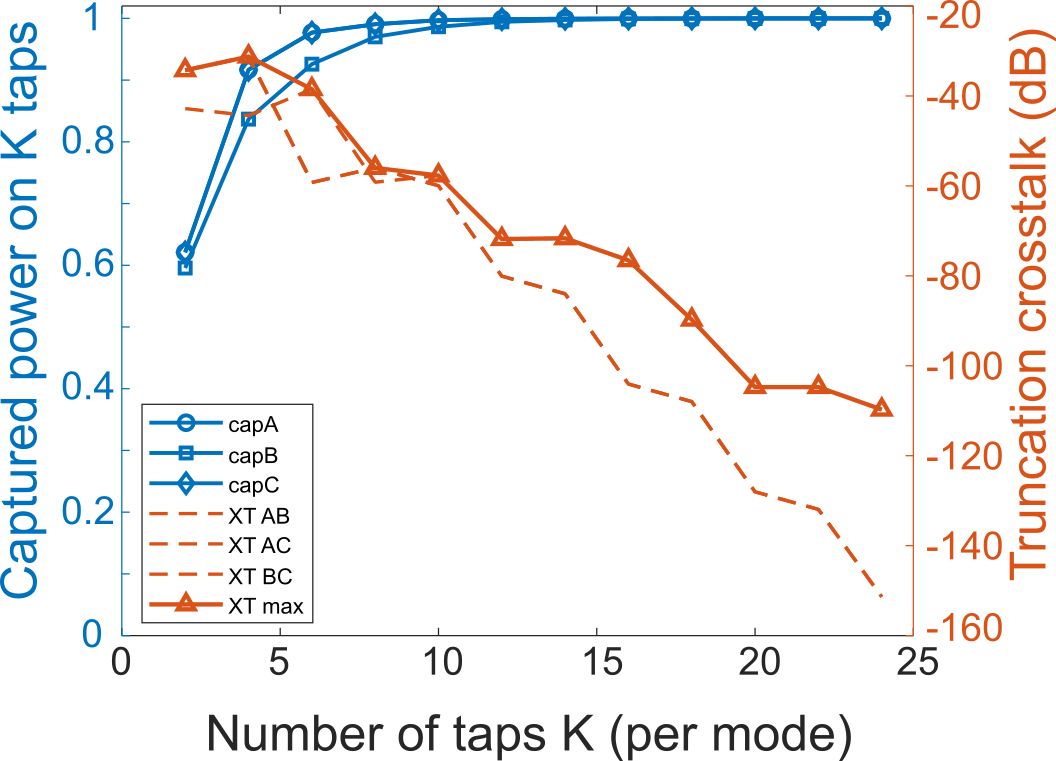}
    \caption{ \textbf{Finite-aperture coherent demultiplexing performance for the $J=4$ device with three in-gap topological modes (A, B, C).} Left axis: captured modal power $\mathrm{cap}_{\mu}(K)$ on the top-$K$ tapped sites for each mode. Right axis: truncation-induced crosstalk between mode pairs, reported as $10\log_{10}(\mathrm{xTalk}_{\mu\nu})$, together with the worst-case value $\mathrm{xTalk}_{\max}$ over all pairs. As $K$ increases, captured power rapidly approaches unity while crosstalk decreases (more negative dB), demonstrating that a compact finite-aperture readout can achieve high collection efficiency with low inter-mode leakage.}
    \label{Fig:tap}
\end{figure}

To experimentally separate the three in-gap topological eigenmodes of the $J=4$ superlattice, we implement a finite-aperture coherent readout that accesses only a limited subset of waveguides around the interface. Because the topological modes are exponentially localized, the modal energy is concentrated on a small number of sites; consequently, a compact on-chip demultiplexing network can be realized with a modest number of taps, without requiring access to the full array. We quantify the feasibility of this approach using two complementary metrics: (i) the \emph{captured power} of each mode on the tapped subset and (ii) the \emph{truncation-induced loss of orthogonality}, which sets the fundamental crosstalk floor of any finite-aperture projector. In the $J=4$ device, three topological modes (A, B, C) appear in the gap, and we evaluate all three modes and their pairwise separability (Fig.~\ref{Fig:tap}).

Let $\mathbf{v}_{A},\mathbf{v}_{B},\mathbf{v}_{C}\in\mathbb{C}^{N}$ denote the (normalized) waveguide-basis eigenvectors of the three topological modes, with components $v_{\mu}(n)$ for $\mu\in\{A,B,C\}$. For a chosen number of taps $K$, we define the tapped set for each mode as the indices of its $K$ largest-magnitude components,
\begin{equation}
\mathcal{S}_{\mu}(K)=\operatorname*{arg\,topK}_{n}\{|v_{\mu}(n)|\},\qquad \mu\in\{A,B,C\}.
\end{equation}
The \emph{captured power fraction}---the fraction of modal energy accessible when only these $K$ sites are tapped---is
\begin{equation}
\mathrm{cap}_{\mu}(K)=\frac{\sum_{n\in \mathcal{S}_{\mu}(K)} |v_{\mu}(n)|^{2}}{\sum_{n=1}^{N} |v_{\mu}(n)|^{2}},\qquad \mu\in\{A,B,C\}.
\label{eq:capABC}
\end{equation}
(For normalized eigenvectors the denominator equals unity, but we retain the explicit form for clarity.)

A finite-aperture readout also modifies the effective inner product because sites outside the tapped set are ignored. Even if the full eigenmodes are orthogonal, $\mathbf{v}_{\mu}^{\dagger}\mathbf{v}_{\nu}=0$ for $\mu\neq\nu$, truncation to a restricted set generally yields non-orthogonal restricted vectors. To model a practical demultiplexer that must operate on a \emph{common} hardware aperture, we define the union of tapped sites
\begin{equation}
\mathcal{S}_{\mathrm{union}}(K)=\mathcal{S}_{A}(K)\cup\mathcal{S}_{B}(K)\cup\mathcal{S}_{C}(K),
\end{equation}
and the corresponding restricted vectors $\mathbf{v}_{\mu,\mathcal{S}}$ obtained by keeping only entries on $\mathcal{S}_{\mathrm{union}}(K)$. We then quantify the truncation-induced crosstalk between any pair of modes $(\mu,\nu)$ by the normalized overlap
\begin{equation}
\mathrm{xTalk}_{\mu\nu}(K)=
\frac{\big|\mathbf{v}_{\mu,\mathcal{S}}^{\dagger}\mathbf{v}_{\nu,\mathcal{S}}\big|^{2}}
{\big(\mathbf{v}_{\mu,\mathcal{S}}^{\dagger}\mathbf{v}_{\mu,\mathcal{S}}\big)\big(\mathbf{v}_{\nu,\mathcal{S}}^{\dagger}\mathbf{v}_{\nu,\mathcal{S}}\big)},
\qquad 
\mathcal{S}=\mathcal{S}_{\mathrm{union}}(K),
\label{eq:xtalkABC}
\end{equation}
which we report in decibels as $10\log_{10}\!\big(\mathrm{xTalk}_{\mu\nu}(K)\big)$. For a single-number design target we also define the worst-case (limiting) crosstalk,
\begin{equation}
\mathrm{xTalk}_{\max}(K)=\max\{\mathrm{xTalk}_{AB}(K),\,\mathrm{xTalk}_{AC}(K),\,\mathrm{xTalk}_{BC}(K)\},
\end{equation}
which represents the intrinsic leakage floor when attempting to realize three matched-filter projectors using only a finite aperture.

Figure~\ref{Fig:tap} shows that the captured fractions $\mathrm{cap}_{A}(K)$, $\mathrm{cap}_{B}(K)$, and $\mathrm{cap}_{C}(K)$ rise rapidly toward unity as $K$ increases, consistent with strong exponential localization: by a modest number of taps, each mode’s energy is already largely contained within the accessed subset. In parallel, the truncation-induced overlaps $\mathrm{xTalk}_{AB}$, $\mathrm{xTalk}_{AC}$, and $\mathrm{xTalk}_{BC}$ decrease monotonically (becoming more negative in dB), indicating progressively improved separability as the readout aperture expands. Importantly, the worst-case crosstalk $\mathrm{xTalk}_{\max}$ improves by many tens of decibels over the explored range, confirming that a compact coherent-combining network can both (i) collect essentially all localized modal weight and (ii) maintain low inter-mode leakage. These trends make mode-selective readout practical with limited hardware complexity, enabling robust interferometric and correlation measurements directly in the topological-mode basis. Based on the trade-off between footprint/routing complexity and separability, a small tap count near the onset of saturation (e.g., $K\sim 5-7$ per mode in this design) already captures the dominant modal weight while providing strong suppression of truncation-induced crosstalk.

\suppheading{S8. Programmable biphoton routing by coherent topological modal superposition}

The multi-band topological superlattice supplies more than a set of
spatially localized interface modes: it provides a discrete,
disorder-protected modal manifold that can serve as a programmable
basis for nonlinear quantum-state generation. In a conventional source
the pump is launched into a single waveguide or a single localized
mode, so the generated biphoton state is fixed by the spatial overlap
between that pump and the signal--idler modes [Eq.~\eqref{eq:selectivity_rule}].
Here we promote the pump to an active control field: by preparing it as
a coherent superposition of several topological modes, the modal
weights themselves select which nonlinear pathways contribute to the
final state, reconfiguring the output without altering the physical
device.

For a superlattice that supports localized topological modes \(M\), the pump input can be written as
\(\psi_{\mathrm{p}}^{\mathrm{in}}\propto \sum_{m=1}^{M} c_m \phi_m^{(p)}\),
where \(\phi_m^{(p)}\) is the \(m\)-th topological pump mode and \(c_m\) is a real-valued modal weight. The total launched pump power is kept fixed by normalizing the coefficient vector, so that changes in the output state originate from the relative modal composition rather than from a trivial variation of pump power. Since the spontaneous four-wave mixing source term is proportional to the square of the pump field, the generated biphoton amplitude depends quadratically on the modal weights. As a result, the source contains both self-products, \(c_m^2 \phi_m^{(p)}\phi_m^{(p)}\) and cross-products, \(c_m c_n \phi_m^{(p)}\phi_n^{(p)}\). These terms contribute coherently, and their relative signs and magnitudes determine whether a given signal--idler channel is enhanced or suppressed.

The biphoton wavefunction obeys the driven two-particle evolution
\(i\partial_z \Psi = H_s\Psi+\Psi H_i^{T}+\gamma\,\mathrm{diag}[A_p^2(z)]\),
where \(H_s\) and \(H_i\) are the signal and idler tight-binding Hamiltonians, \(A_p(z)\) is the propagated pump field and \(\gamma\) is the nonlinear coefficient. Projecting the final biphoton state onto the topological signal and idler modes gives the eigenmode-basis amplitude
\(\mathcal{A}_{\mu_s\mu_i}=\langle \phi_{\mu_s}^{(s)}|\Psi(L)|\phi_{\mu_i}^{(i)}\rangle\),
with population
\(P_{\mu_s\mu_i}=|\mathcal{A}_{\mu_s\mu_i}|^2\).
Equivalently, the amplitude in each biphoton block can be viewed as a coherent sum over all pump-mode products,
\[
\mathcal{A}_{\mu_s\mu_i}
\propto
\sum_{m\leq n}\eta_{mn}c_m c_n\,\mathcal{G}_{\mu_s\mu_i}^{(mn)},
\]
where \(\eta_{mn}=1\) for \(m=n\) and \(\eta_{mn}=2\) for \(m\neq n\), and \(\mathcal{G}_{\mu_s\mu_i}^{(mn)}\) represents the nonlinear propagated response associated with the pump-mode product \(\phi_m^{(p)}\phi_n^{(p)}\). This expression clarifies the physical origin of the control mechanism: the coefficient vector \(\mathbf{c}\) does not simply scale the output intensity, but reshapes the coherent interference among multiple nonlinear generation pathways. The topological mode basis therefore provides a compact and robust platform in which the biphoton population can be redistributed by tuning the pump superposition.

\begin{figure}[H]
    \centering
    \includegraphics[width=\textwidth]{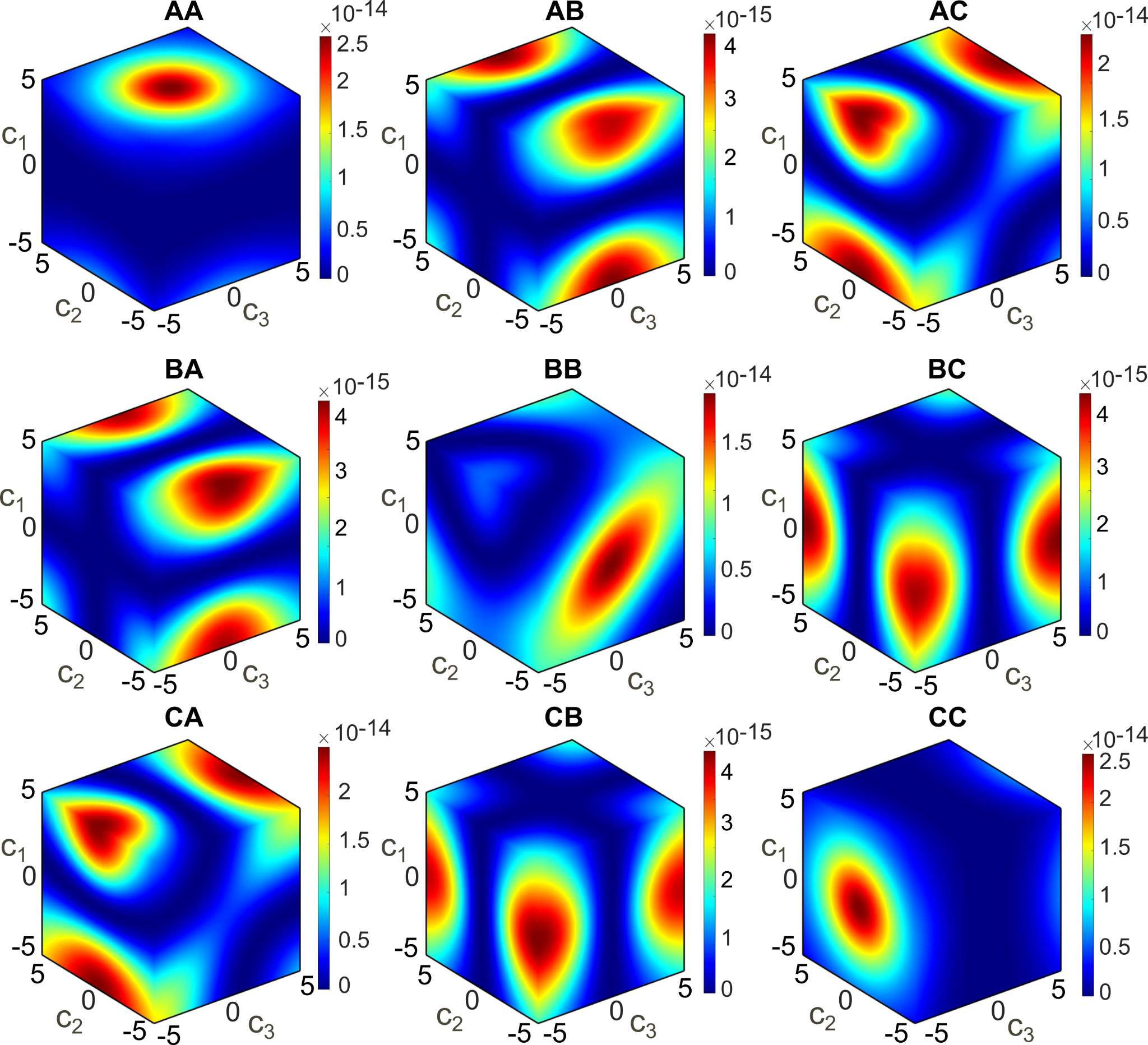}
    \caption{\textbf{Coefficient-space control of biphoton populations
    ($J=4$).} Biphoton eigenmode populations $P_{\mu_s\mu_i}$ for the
    nine signal--idler blocks $\{AA,\ldots,CC\}$ versus the real
    pump-superposition coefficients $(c_1,c_2,c_3)$, with the pump
    prepared as $\psi_{\mathrm{p}}^{\mathrm{in}}\propto
    c_1\phi_A+c_2\phi_B+c_3\phi_C$ at fixed total power. Distinct high-population regions across the
    panels show that the generated state can be redistributed among
    topological signal--idler channels by tuning the relative modal
    weights alone, through coherent interference of the self- and
    cross-products of the pump modes.}
    \label{fig:J4_blocks}
\end{figure}

Figure~\ref{fig:J4_blocks} visualizes this mechanism for the \(J=4\) superlattice, where three localized topological modes \(A\)--\(C\) form a three-dimensional pump basis and a \(3\times3\) signal--idler output space. The pump state is written as \(\psi_{\mathrm{p}}^{\mathrm{in}}\propto c_1\phi_A+c_2\phi_B+c_3\phi_C\), and the nine panels show the calculated populations of all biphoton blocks \(\{AA,AB,AC,BA,BB,BC,CA,CB,CC\}\) over the coefficient space \(c_1,c_2,c_3\in[-5,5]\). The different panels exhibit distinct high-population regions, showing that each signal--idler channel responds differently to the same pump-control space. This behavior is a direct consequence of the block-dependent response tensors \(\mathcal{G}_{\mu_s\mu_i}^{(mn)}\), which cause the self- and cross-pump contributions to interfere differently for each output channel. Thus, the coefficient space acts as a nonlinear control landscape for biphoton routing.

\begin{figure}[H]
    \centering
    \includegraphics[width=\textwidth]{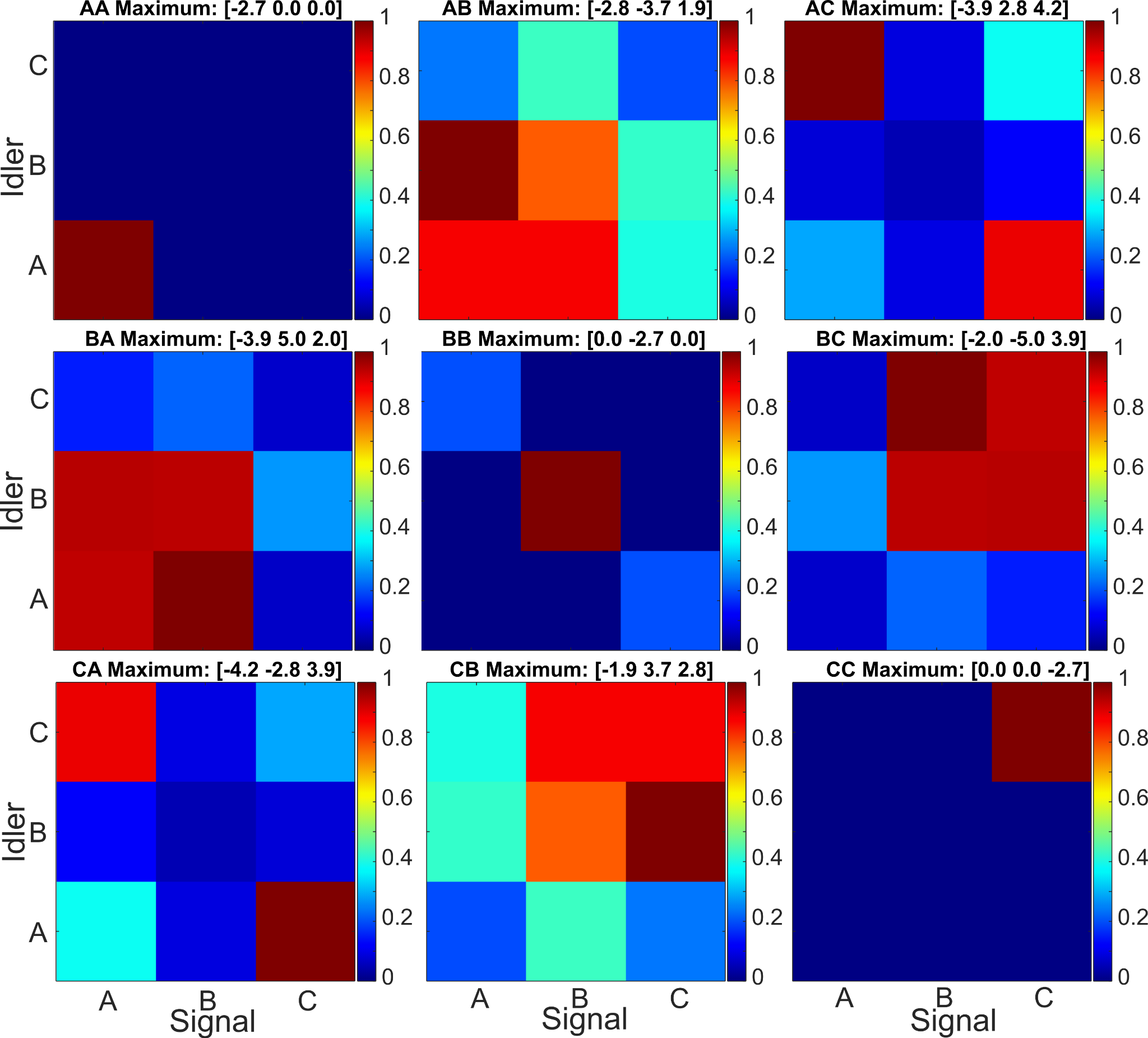}
    \caption{\textbf{Selective optimization of biphoton populations
    ($J=4$).} Normalized population matrices $P_{\mu_s\mu_i}$ for
    optimized real pump superpositions of the three topological modes
    $A$--$C$; the coefficient vector $\mathbf{c}$ maximizing the
    indicated block is given above each panel. Axes label the signal
    (horizontal) and idler (vertical) topological modes. Both diagonal
    and off-diagonal blocks are individually addressable, demonstrating
    selective routing of photon-pair correlations within the
    topological manifold.}
    \label{fig:J4_opt}
\end{figure}

The optimized examples in Fig.~\ref{fig:J4_opt} show that this control landscape can be used to selectively enhance specific biphoton eigenmode populations in the \(J=4\) system. For each target block, the pump coefficients are chosen to maximize the desired signal--idler channel after propagation. The resulting matrices demonstrate that both diagonal and off-diagonal biphoton correlations can be addressed through coherent modal tailoring. This is important because off-diagonal blocks correspond to photon pairs occupying different topological modes, which are essential for generating structured high-dimensional correlations rather than only same-mode pair emission.

\begin{figure}[H]
    \centering
    \includegraphics[width=1\textwidth]{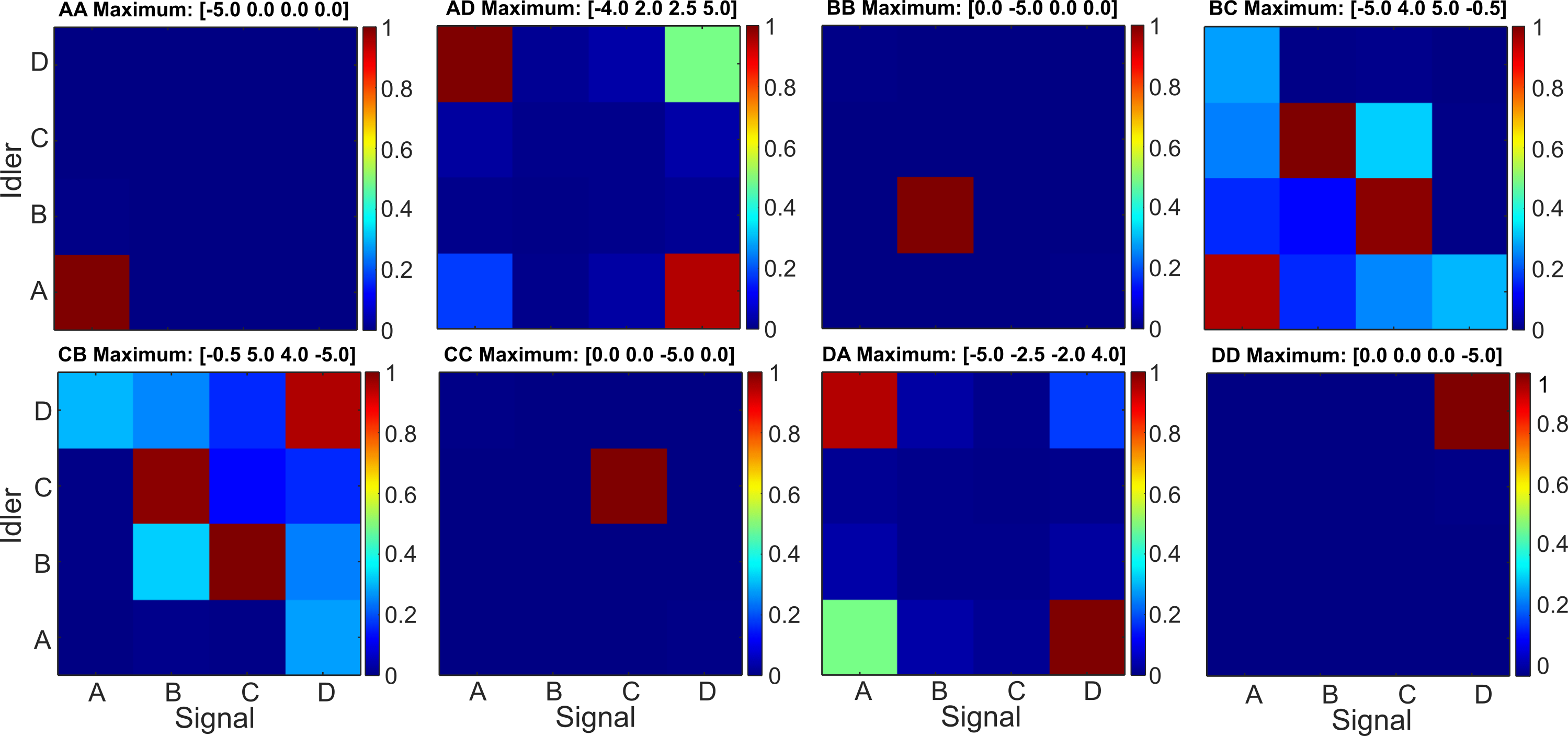}
    \caption{
    \textbf{Selective optimization of biphoton eigenmode populations in the $J=5$ topological superlattice.}
    Normalized biphoton eigenmode population matrices obtained for selected optimized pump superpositions of the four localized topological modes $A$--$D$. 
    In each panel, the pump input is prepared as a coherent modal superposition $\psi_{\mathrm{p}}^{\mathrm{in}} \propto c_1\phi_A+c_2\phi_B+c_3\phi_C+c_4\phi_D$, where the coefficient vector shown above each panel gives the optimized real-valued pump weights that maximize the indicated biphoton Eigenmode population block. 
    The horizontal and vertical axes denote the signal and idler topological modes, respectively, and the color scale represents the normalized biphoton population in the eigenmode basis, $P_{\mu_s\mu_i}=\left|\langle \phi_{\mu_s}^{(s)}|\Psi(L)|\phi_{\mu_i}^{(i)}\rangle\right|^2$.
    The results show that coherent tailoring of the pump weights enables selective routing of the generated photon-pair state into distinct topological mode-pair channels, providing a programmable mechanism for controlling high-dimensional biphoton correlations in the topological basis.
    }
    \label{J5-BPM}
\end{figure}

The same principle extends naturally to larger superlattices. In the \(J=5\) case, shown in Fig.~\ref{J5-BPM}, the pump is expanded over four localized topological modes \(A\)--\(D\), \(\psi_{\mathrm{p}}^{\mathrm{in}}\propto c_1\phi_A+c_2\phi_B+c_3\phi_C+c_4\phi_D\). The corresponding biphoton population space contains \(4\times4\) signal--idler channels. The additional pump mode increases the number of coherent self- and cross-generation pathways, thereby enlarging the accessible control space. The optimized matrices show that selected biphoton blocks can still be preferentially enhanced, confirming that the method remains effective beyond the three-mode case.

\begin{figure}[H]
    \centering
    \includegraphics[width=1\textwidth]{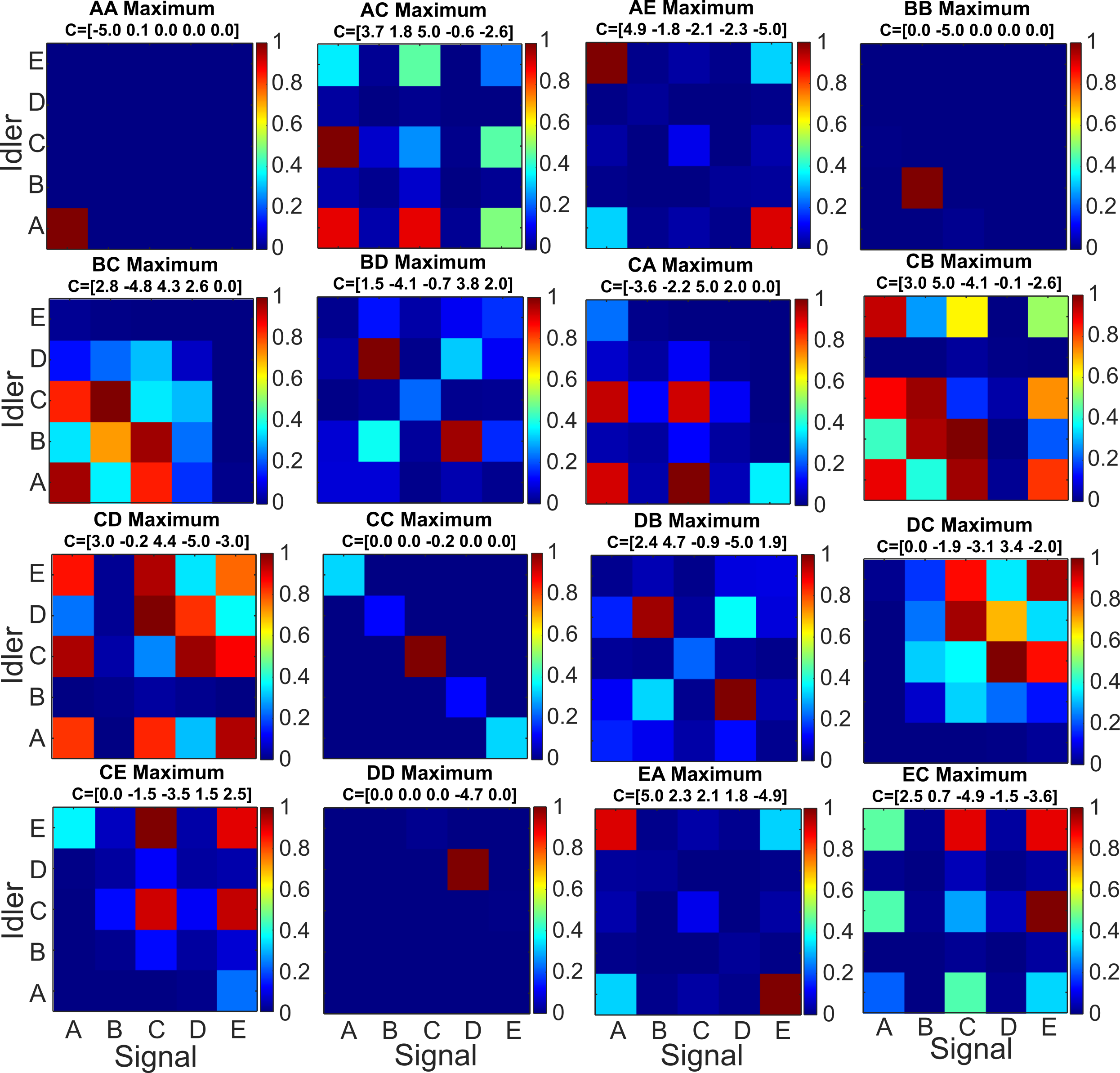}
    \caption{
    \textbf{Selective optimization of biphoton eigenmode populations in the $J=6$ topological superlattice.}
    Normalized biphoton eigenmode population matrices obtained for selected optimized pump superpositions of the five localized topological modes $A$--$E$. 
    In each panel, the pump input is prepared as a coherent modal superposition $\psi_{\mathrm{p}}^{\mathrm{in}} \propto c_1\phi_A+c_2\phi_B+c_3\phi_C+c_4\phi_D+c_5\phi_E$, where the coefficient vector shown above each panel gives the optimized real-valued pump weights that maximize the indicated signal--idler block. 
    The horizontal and vertical axes denote the signal and idler topological modes, respectively, and the color scale represents the normalized biphoton population in the eigenmode basis, $P_{\mu_s\mu_i}=\left|\langle \phi_{\mu_s}^{(s)}|\Psi(L)|\phi_{\mu_i}^{(i)}\rangle\right|^2$.
    These results demonstrate that coherent tailoring of the five-mode pump superposition enables selective enhancement of distinct photon-pair channels within the topological-mode manifold, providing a programmable route for controlling high-dimensional biphoton correlations in the $J=6$ superlattice.
    }
    \label{J6-BPM}
\end{figure}

Figure~\ref{J6-BPM} further demonstrates the scalability of the approach for the \(J=6\) superlattice, where five localized topological modes \(A\)--\(E\) define a \(5\times5\) biphoton-mode manifold. In this higher-dimensional setting, the pump state \(\psi_{\mathrm{p}}^{\mathrm{in}}\propto c_1\phi_A+c_2\phi_B+c_3\phi_C+c_4\phi_D+c_5\phi_E\) generates an even larger set of interfering nonlinear pathways. The optimized population matrices show that selective enhancement of different target channels remains achievable, indicating that coherent topological modal superposition provides a scalable strategy for engineering high-dimensional biphoton states.

Together these results establish a programmable framework for nonlinear
quantum-state control in topological photonic lattices. The localized
topological modes furnish a robust, discrete basis, and the real pump
coefficients $\mathbf{c}$ form a control vector that sets the
interference among nonlinear generation pathways. Routing the biphoton state among
selected signal--idler channels then requires only reprogramming the
pump superposition---no change to the fabricated lattice---a capability
well matched to integrated quantum photonics, where reconfigurable
access to high-dimensional entangled states underpins
multidimensional quantum interference and on-chip state synthesis.

\suppheading{S9. Quantum metrics calculation}

To quantify the effect of simulation and fabrication-induced disorder on both the dimensionality and stability of the generated biphoton state, we analyze each disorder realization using two complementary figures of merit: an entropy-based effective Schmidt dimensionality and the state fidelity with respect to the disorder-free output. This procedure provides a realization-resolved comparison between the ideal biphoton wavefunction and the corresponding disordered state at the output plane.

For a given disorder realization, the output two-photon state is represented in the discrete spatial basis as
\begin{equation}
    |\Psi^{(\mathrm{dis})}\rangle
    =
    \sum_{m,n}\psi^{(\mathrm{dis})}_{mn}
    |m_s\rangle |n_i\rangle ,
    \qquad
    \sum_{m,n}\left|\psi^{(\mathrm{dis})}_{mn}\right|^2=1,
\end{equation}
where $m$ and $n$ label the output signal and idler waveguides, respectively. The biphoton amplitude matrix
$\boldsymbol{\psi}^{(\mathrm{dis})}$ is then decomposed by singular-value decomposition,
\begin{equation}
    \boldsymbol{\psi}^{(\mathrm{dis})}
    =
    U S V^\dagger ,
    \qquad
    S=\mathrm{diag}(s_1,s_2,\ldots),
\end{equation}
where $s_k\geq 0$ are the Schmidt singular values. The corresponding normalized Schmidt weights are
\begin{equation}
    p_k
    =
    \frac{s_k^2}{\sum_j s_j^2},
    \qquad
    \sum_k p_k=1 .
\end{equation}
These weights define the Schmidt-mode probability distribution of the biphoton state. We therefore introduce the entropy-equivalent Schmidt dimensionality
\begin{equation}
    K_{\mathrm{eff}}
    =
    \exp\!\left[-\sum_k p_k \ln p_k\right],
\end{equation}
which gives the effective number of coherently occupied Schmidt modes \cite{Law2000continuous}. This quantity equals unity for a separable two-photon state and approaches the actual number of equally populated Schmidt modes for a maximally distributed $D$-mode state, for which $K_{\mathrm{eff}}=D$. It therefore provides a compact measure of the disorder-induced redistribution of biphoton population among the available topological modes \cite{Peschel2012entanglement}.

The stability of the generated state is quantified by the pure-state fidelity between the ideal disorder-free biphoton amplitude $\psi^{(\mathrm{ideal})}_{mn}$ and the disordered amplitude $\psi^{(\mathrm{dis})}_{mn}$,
\begin{equation}
    F_{\psi}
    =
    \left|
    \langle \Psi^{(\mathrm{ideal})}|\Psi^{(\mathrm{dis})}\rangle
    \right|^2
    =
    \left|
    \sum_{m,n}
    \left(\psi^{(\mathrm{ideal})}_{mn}\right)^{*}
    \psi^{(\mathrm{dis})}_{mn}
    \right|^2 .
\end{equation}
Here, both states are normalized to unity before evaluating the overlap. Thus, $F_{\psi}=1$ corresponds to an unchanged output biphoton wavefunction, whereas reductions in $F_{\psi}$ directly quantify disorder-induced changes in the coherent two-photon state. For each disorder strength, the reported values of $K_{\mathrm{eff}}$ and $F_{\psi}$ are obtained over the ensemble of disorder realizations, allowing the mean response and the realization-to-realization fluctuations to be assessed.

In the experiment, the measurement gives access to joint spatial intensity correlations, i.e., coincidence probabilities, rather than the full complex biphoton amplitude. Consequently, the phase of $\psi_{mn}$ is not experimentally reconstructed, and a complete quantum-state fidelity cannot be directly extracted from the measured correlations alone. We therefore compare experiment and theory using the classical Bhattacharyya fidelity between the normalized measured and simulated joint probability distributions,
\begin{equation}
    P_{mn}
    =
    \frac{C^{(\mathrm{exp})}_{mn}}
    {\sum_{m,n} C^{(\mathrm{exp})}_{mn}},
    \qquad
    Q_{mn}
    =
    \frac{C^{(\mathrm{th})}_{mn}}
    {\sum_{m,n} C^{(\mathrm{th})}_{mn}},
\end{equation}
where $C^{(\mathrm{exp})}_{mn}$ and $C^{(\mathrm{th})}_{mn}$ denote the measured and calculated coincidence maps, respectively. The intensity-level fidelity is then defined as
\begin{equation}
    F_{\mathrm{I}}
    =
    \left(
    \sum_{m,n}
    \sqrt{P_{mn}Q_{mn}}
    \right)^2 .
\end{equation}
This phase-insensitive metric quantifies the agreement between measured and simulated biphoton population distributions. Equivalently, it represents the maximum pure-state overlap compatible with the measured intensities when the relative phases are not known. We therefore use $F_{\mathrm{I}}$ as an experimental measure of spatial-correlation agreement, while $F_{\psi}$ is reserved for the full complex-amplitude fidelity available in simulation.

\suppheading{S10. Targeted synthesis of prescribed Schmidt dimensionality}

A key objective in programmable quantum photonic lattices is to move beyond the passive generation of entanglement and toward the deterministic synthesis of biphoton states with a prescribed effective dimensionality. In this work, the Schmidt number is treated as a targetable design parameter controlled by coherent excitation of the topological pump modes. For each lattice configuration \(J\), the pump field is written as a superposition of the available topological modes with real coefficients \(\{c_m\}\). The coefficient space is then searched to identify the five launch conditions that generate biphoton states closest to a selected target Schmidt number. These candidate solutions are first obtained using the precomputed biphoton response matrices and are subsequently verified by direct propagation of the coupled pump and biphoton dynamics. This two-step procedure provides both computational efficiency and an independent consistency check, ensuring that the identified states are not numerical artifacts of the reduced search space.

For the \(J=4\) configuration, three topological pump modes are used to target 
\(K_{\mathrm{target}}=3.4100\). The five closest coefficient sets are listed in Table~\ref{tab:J4}. All selected solutions reproduce the target value with high accuracy, giving Schmidt numbers in the narrow interval \(K=3.4100\)--\(3.4101\). The two closest solutions, \((-2.5,\,4.3,\,-3.1)\) and \((-3.1,\,4.3,\,-2.5)\), reach the target with an absolute error of only \(4.8852\times10^{-6}\). The agreement between the precomputed result and the direct-propagation check is identical to the displayed precision, confirming that the response-matrix method accurately captures the full nonlinear biphoton generation dynamics.

\begin{figure}[H]
    \centering
    \includegraphics[width=1\textwidth]{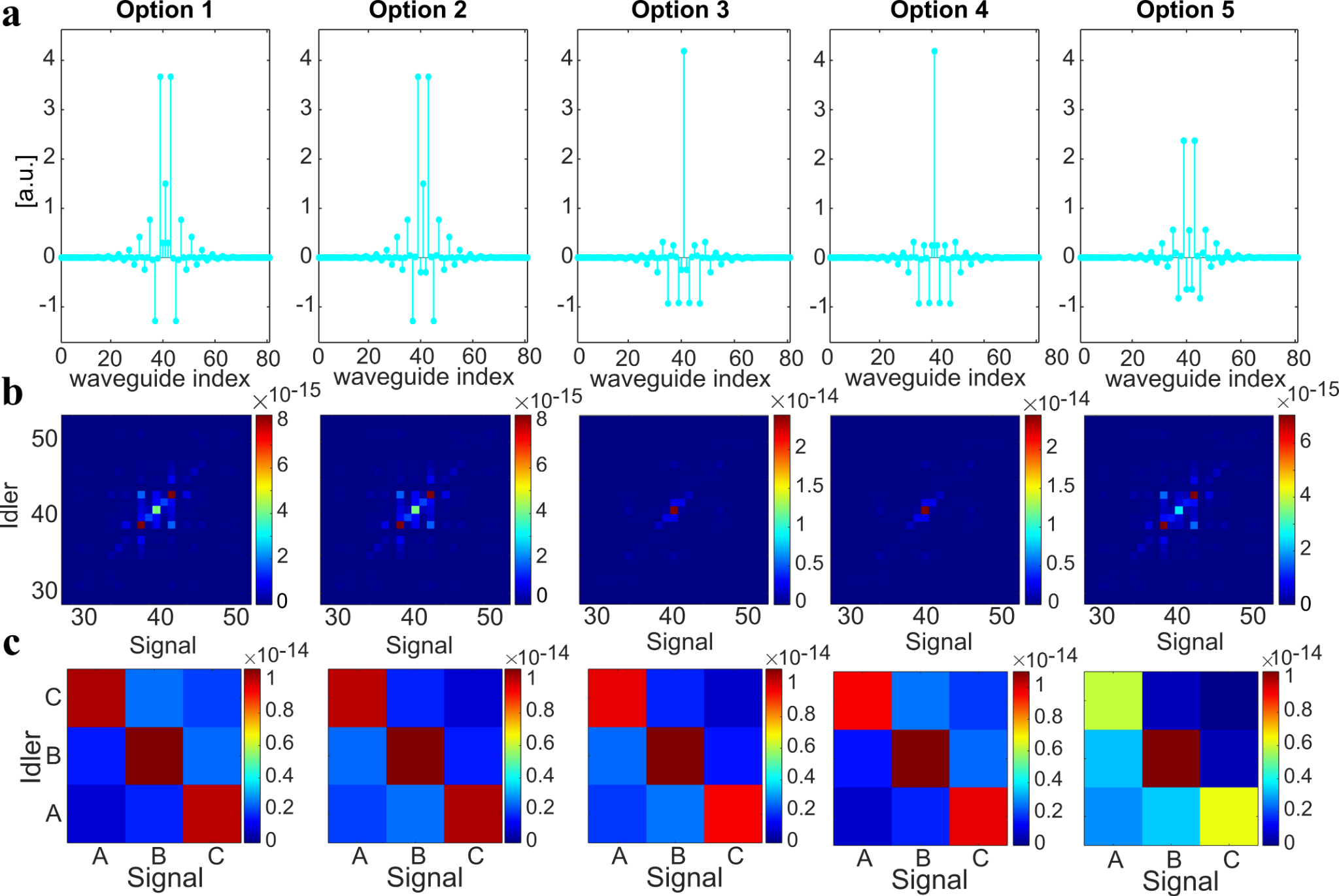}
\caption{\textbf{Inverse-designed generation of biphoton states with prescribed Schmidt dimensionality in the \(J=4\) topological superlattice.}
Five optimized pump configurations are shown, each targeting \(K_{\mathrm{target}}=3.41\). The five coefficient sets are those whose calculated Schmidt numbers lie closest to the target value. 
\textbf{a,} Real-space pump excitation profiles generated by coherent superpositions of the three localized topological pump modes. 
\textbf{b,} Corresponding signal--idler correlation distributions in the waveguide basis, showing the spatial structure of the generated biphoton correlations near the central topological region. 
\textbf{c,} Biphoton population matrices projected onto the three-mode topological subspace, with modes labelled \(A\)–\(C\) for both signal and idler photons. 
Although the five optimized inputs yield nearly identical Schmidt dimensionality, they populate distinct topological mode-pair channels and generate different correlation patterns. These results show that coherent pump programming can prescribe the effective dimensionality of entanglement while independently shaping the underlying modal correlation pathway.}
\label{fig:target_schmidt_j4}
\end{figure}

The physical structure of these five \(J=4\) solutions is shown in Fig.~\ref{fig:target_schmidt_j4}. Figure~\ref{fig:target_schmidt_j4}a shows that the optimized pump profiles are strongly localized around the central region of the array, but differ in their relative amplitudes and phases across the participating waveguides. These differences lead to distinct signal--idler joint intensity patterns in the site basis, as shown in Fig.~\ref{fig:target_schmidt_j4}b. Although all five cases yield nearly the same Schmidt number, their spatial correlation maps are not identical, indicating that the same entanglement dimensionality can be reached through different biphoton wavefunctions. This distinction becomes clearer in Fig.~\ref{fig:target_schmidt_j4}c, where the generated states are projected onto the three-mode topological subspace. The population matrices show different distributions among the A, B, and C signal--idler channels, demonstrating that a fixed Schmidt number does not uniquely specify the modal composition of the biphoton state. Instead, it defines a family of states with comparable entanglement entropy but different microscopic correlation pathways.

\begin{table*}[t]
\centering

\caption{\textbf{Five closest J=4 coefficient options to the target Schmidt number.}}
\label{tab:J4}

\renewcommand{\arraystretch}{1.45}
\setlength{\tabcolsep}{6pt}

\resizebox{\textwidth}{!}{%
\Large
\begin{tabular}{c c c c c c c}
\toprule
\textbf{Option} &
$\mathbf{c_1}$ &
$\mathbf{c_2}$ &
$\mathbf{c_3}$ &
\textbf{\begin{tabular}{c}K Direct\\Check\end{tabular}} &
\textbf{\begin{tabular}{c}Error Direct\\Check\end{tabular}} &
\textbf{\begin{tabular}{c}Target\\K\end{tabular}} \\
\midrule

1 & -2.5 &  4.3 & -3.1 & 3.4100 & $4.8852\times10^{-6}$ & 3.4100 \\

2 & -3.1 &  4.3 & -2.5 & 3.4100 & $4.8852\times10^{-6}$ & 3.4100 \\

3 & -2.4 & -3.5 & -1.9 & 3.4101 & $6.5968\times10^{-5}$ & 3.4100 \\

4 & -1.9 & -3.5 & -2.4 & 3.4101 & $6.5968\times10^{-5}$ & 3.4100 \\

5 & -2.2 &  3.0 & -0.9 & 3.4101 & $7.4978\times10^{-5}$ & 3.4100 \\

\bottomrule
\end{tabular}%
}
\end{table*}

For the \(J=5\) configuration, the accessible topological pump space increases to four modes, allowing the target Schmidt number to be raised to 
\(K_{\mathrm{target}}=4.3700\). As shown in Table~\ref{tab:J5}, all five optimized coefficient sets reproduce the target value to four decimal places. The precomputed errors are zero at the numerical precision reported in the table, while the direct-propagation verification gives residual discrepancies only at the \(10^{-13}\) level. Representative optimized coefficient vectors include 
\((-1.9973,\,-4.0003,\,-4.5112,\,1.0027)\), 
\((-2.0001,\,5.0011,\,2.0006,\,-4.0081)\), and 
\((-0.50343,\,1.4995,\,3.5029,\,-0.50447)\). These markedly different pump superpositions all converge to the same target Schmidt dimensionality, highlighting the non-unique relationship between the pump-launch vector and the final entanglement entropy.

\begin{figure}[H]
    \centering
    \includegraphics[width=1\textwidth]{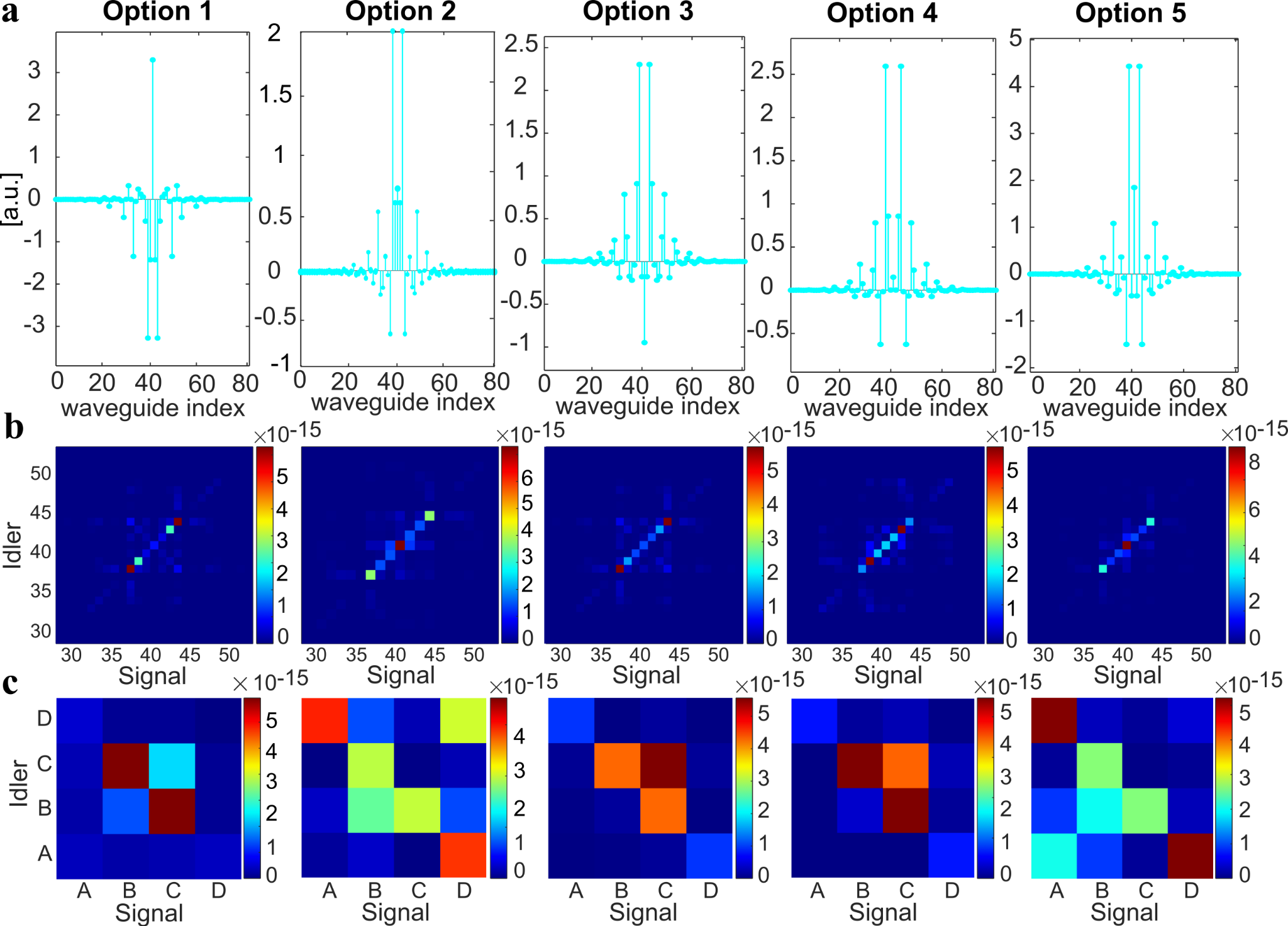}
\caption{\textbf{Inverse-designed generation of biphoton states with prescribed Schmidt dimensionality in the \(J=5\) topological superlattice.}
Five optimized pump configurations are shown for the \(J=5\) lattice, each targeting \(K_{\mathrm{target}}=4.37\). The five coefficient sets are those whose calculated Schmidt numbers lie closest to the target value. 
\textbf{a,} Real-space pump excitation profiles generated by coherent superpositions of the selected topological pump modes. 
\textbf{b,} Corresponding signal--idler correlation distributions in the waveguide basis, showing the spatial structure of the generated biphoton correlations localized around the central topological region. 
\textbf{c,} Biphoton population matrices projected onto the four-mode topological subspace, with modes labelled \(A\)–\(D\) for both signal and idler photons. 
Although the five optimized inputs yield nearly identical Schmidt dimensionality, they populate distinct topological mode-pair channels and generate different correlation patterns. These results show that, in the \(J=5\) superlattice, coherent pump programming can prescribe the effective dimensionality of entanglement while independently shaping the underlying modal correlation pathway.}
\label{fig:target_schmidt_J5}
\end{figure}

The corresponding field and correlation structures are presented in Fig.~\ref{fig:target_schmidt_J5}. The pump profiles in Fig.~\ref{fig:target_schmidt_J5}(A) show that the additional topological degree of freedom provides more flexibility in shaping the central excitation pattern. The generated signal--idler intensity maps in Fig.~\ref{fig:target_schmidt_J5}(B) remain concentrated near the topological region, but the diagonal and near-diagonal features vary from one option to another. This behavior indicates that the biphoton pairs remain spatially correlated while the detailed distribution of the correlations can be reconfigured by changing the pump-mode coefficients. In the projected four-mode basis, Fig.~\ref{fig:target_schmidt_J5}(C) reveals a broader redistribution of population over the A--D modal channels compared with the \(J=4\) case. Thus, increasing \(J\) not only increases the reachable Schmidt dimensionality, but also expands the number of modal pathways through which that dimensionality can be realized.

\begin{table*}[t]
\centering
\caption{\textbf{Five closest J=5 coefficient options to the target Schmidt number.}}
\label{tab:J5}

\renewcommand{\arraystretch}{1.45}
\setlength{\tabcolsep}{6pt}

\resizebox{\textwidth}{!}{%
\Large
\begin{tabular}{c c c c c c c c}
\toprule
\textbf{Option} &
$\mathbf{c_1}$ &
$\mathbf{c_2}$ &
$\mathbf{c_3}$ &
$\mathbf{c_4}$ &
\textbf{\begin{tabular}{c}K Direct\\Check\end{tabular}} &
\textbf{\begin{tabular}{c}Error Direct\\Check\end{tabular}} &
\textbf{\begin{tabular}{c}Target\\K\end{tabular}} \\
\midrule

1 & -1.9973     & -4.0003 & -4.5112 &  1.0027   & 4.3700 & $4.7873\times10^{-13}$ & 4.3700 \\

2 & -2.0001     &  5.0011 &  2.0006 & -4.0081   & 4.3700 & $1.4033\times10^{-13}$ & 4.3700 \\

3 & -0.50343    &  1.4995 &  3.5029 & -0.50447  & 4.3700 & $2.7534\times10^{-14}$ & 4.3700 \\

4 &  0.00057933 & -2.0029 &  3.5046 & -1.0042   & 4.3700 & $1.7408\times10^{-13}$ & 4.3700 \\

5 & -4.0023     &  5      &  2.001  & -2.5109   & 4.3700 & $2.5757\times10^{-13}$ & 4.3700 \\

\bottomrule
\end{tabular}%
}
\end{table*}

For the \(J=6\) configuration, the topological subspace expands further to five participating pump modes. The target Schmidt number is increased to 
\(K_{\mathrm{target}}=5.4400\), and the five closest coefficient sets are summarized in Table~\ref{tab:J6}. All selected options reproduce the target value to the reported precision in the precomputed search, while the direct-propagation errors remain on the order of \(10^{-12}\). The presence of both dominant and nearly vanishing coefficients shows that the target state can be synthesized through highly nontrivial interference among the five topological pump components.

\begin{figure}[H]
    \centering
    \includegraphics[width=1\textwidth]{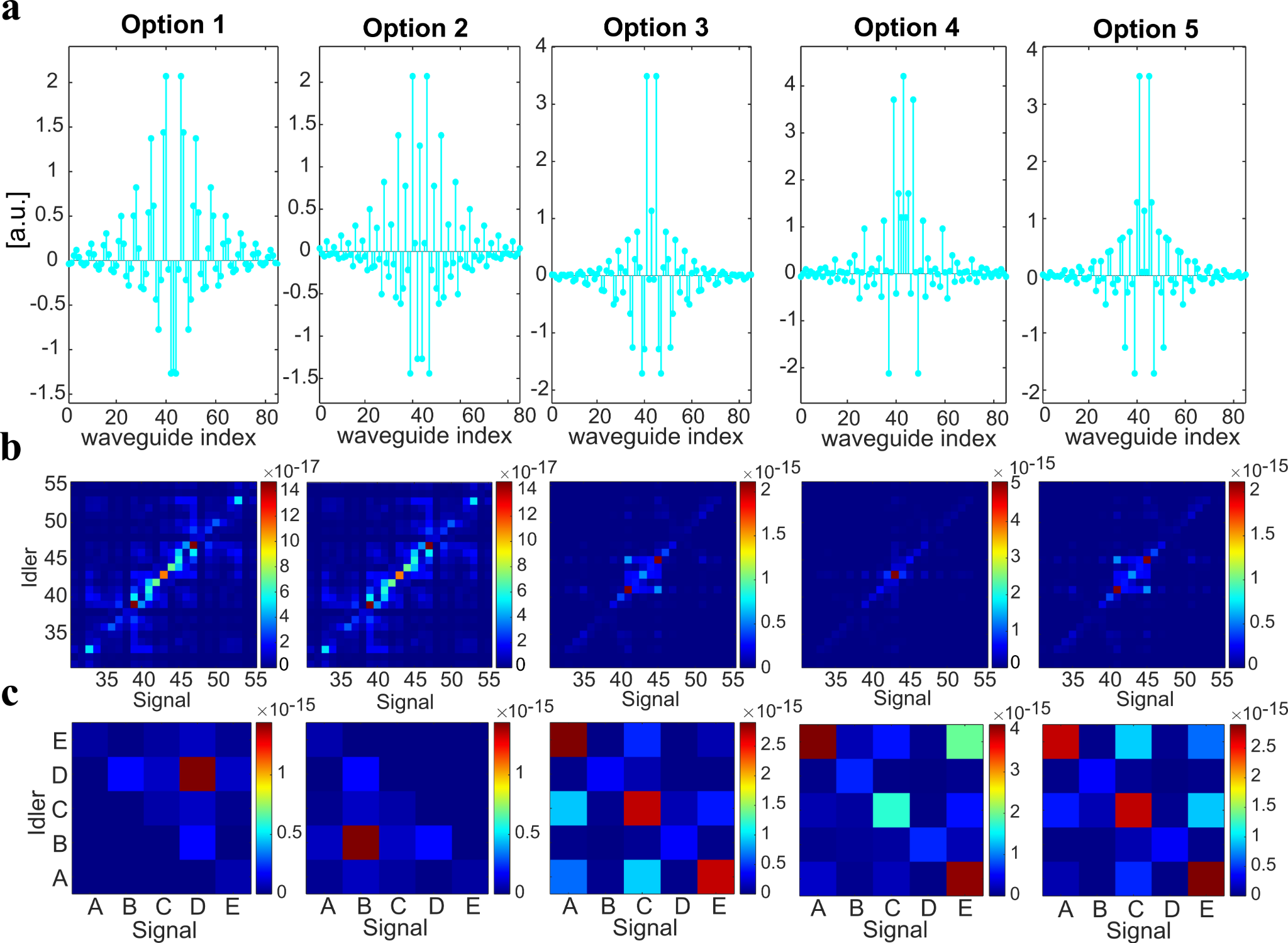}
\caption{\textbf{Inverse-designed generation of biphoton states with prescribed Schmidt dimensionality in the \(J=6\) topological superlattice.}
Five optimized pump configurations are shown for the \(J=6\) lattice, each targeting \(K_{\mathrm{target}}=5.44\). The five coefficient sets are those whose calculated Schmidt numbers lie closest to the target value. 
\textbf{a,} Real-space pump excitation profiles generated by coherent superpositions of the selected topological pump modes. 
\textbf{b,} Corresponding signal--idler correlation distributions in the waveguide basis, showing the spatial structure of the generated biphoton correlations localized around the central topological region. 
\textbf{c,} Biphoton population matrices projected onto the five-mode topological subspace, with modes labelled \(A\)–\(E\) for both signal and idler photons. 
Although the five optimized inputs yield nearly identical Schmidt dimensionality, they populate distinct topological mode-pair channels and produce different correlation patterns. These results show that, in the \(J=6\) superlattice, coherent pump programming can prescribe the effective dimensionality of entanglement while independently shaping the underlying modal correlation pathway.}
\label{fig:target_schmidt_J6}
\end{figure}

Figure~\ref{fig:target_schmidt_J6} visualizes the corresponding \(J=6\) biphoton states. The launch profiles in Fig.~\ref{fig:target_schmidt_J6}(A) display richer spatial structure than in the lower-\(J\) cases, reflecting the enlarged topological modal basis. The site-basis joint intensity distributions in Fig.~\ref{fig:target_schmidt_J6}(B) show that some optimized solutions produce relatively compact correlations near the array center, whereas others distribute the biphoton probability more broadly along the signal--idler diagonal. The projected five-mode population maps in Fig.~\ref{fig:target_schmidt_J6}(C) further demonstrate that the same target Schmidt number can be associated with substantially different modal architectures. Some options concentrate population in a small number of dominant modal-pair channels, while others distribute the population over several signal--idler combinations. This behavior confirms that the Schmidt number constrains the effective entanglement dimensionality but does not uniquely determine the modal state itself.

\begin{table*}[t]
\centering
\caption{\textbf{Five closest J=6 coefficient options to the target Schmidt number.}}
\label{tab:J6}

\renewcommand{\arraystretch}{1.45}
\setlength{\tabcolsep}{0.1pt}

\resizebox{\textwidth}{!}{%
\Large
\begin{tabular}{c c c c c c c c c}
\toprule
\textbf{Option} &
$\mathbf{c_1}$ &
$\mathbf{c_2}$ &
$\mathbf{c_3}$ &
$\mathbf{c_4}$ &
$\mathbf{c_5}$ &
\textbf{\begin{tabular}{c}K Direct\\Check\end{tabular}} &
\textbf{\begin{tabular}{c}Error Direct\\Check\end{tabular}} &
\textbf{\begin{tabular}{c}Target\\K\end{tabular}} \\
\midrule

1 &  0.000314135 & -1.00016 & -0.99969 &  5       &  1.00126     & 5.4400 & $1.6671\times10^{-12}$ & 5.4400 \\

2 & -1.00059     & -5       &  1.00021 &  1.00008 &  0.000242819 & 5.4400 & $1.7151\times10^{-12}$ & 5.4400 \\

3 & -2.99566     &  0.998931&  5       & -2.00541 & -2.00675     & 5.4400 & $1.0276\times10^{-12}$ & 5.4400 \\

4 & -3.00887     &  4.00304 & -4.00275 &  1.99795 & -5           & 5.4400 & $1.2177\times10^{-12}$ & 5.4400 \\

5 & -2.00675     & -2.00541 &  5       &  0.998931& -2.99566     & 5.4400 & $1.0303\times10^{-12}$ & 5.4400 \\

\bottomrule
\end{tabular}%
}
\end{table*}

Taken together, Tables~\ref{tab:J4}--\ref{tab:J6} and Figs.~\ref{fig:target_schmidt_j4}--\ref{fig:target_schmidt_J6} establish a systematic protocol for engineering high-dimensional biphoton entanglement in topological waveguide arrays. As \(J\) increases from 4 to 6, the accessible topological modal subspace expands from three to five modes, enabling target Schmidt numbers to increase from \(K\simeq3.41\) to \(K\simeq4.37\) and \(K\simeq5.44\). Across all three configurations, the optimized coefficients obtained from the response-matrix search are validated by direct propagation with excellent agreement, ranging from errors below \(10^{-4}\) in the coarser \(J=4\) sweep to \(10^{-13}\)--\(10^{-12}\) in the higher-dimensional cases. The figures further show that this control is not limited to matching a scalar entanglement metric: the underlying pump profiles, real-space biphoton correlations, and topological-mode population matrices can all be reshaped while preserving the desired Schmidt dimensionality. These results demonstrate a programmable route to families of topological biphoton states with prescribed entanglement dimensionality and independently tunable modal correlation structure.

To assess the resilience of the targeted synthesis protocol, we next introduce Gaussian normal disorder into the coupling coefficients of the superlattice and monitor the generated biphoton state as the disorder strength is increased. This analysis compares two physically distinct excitation strategies. The first is the conventional topological-superlattice excitation, in which the pump is launched into the central waveguide. The second is the optimized coherent pump superposition identified above for synthesizing a prescribed Schmidt number. In both cases, the input condition is fixed after its definition in the ideal lattice and is not re-optimized for the disordered structures.

\begin{figure}[H]
    \centering
    \includegraphics[width=1\textwidth]{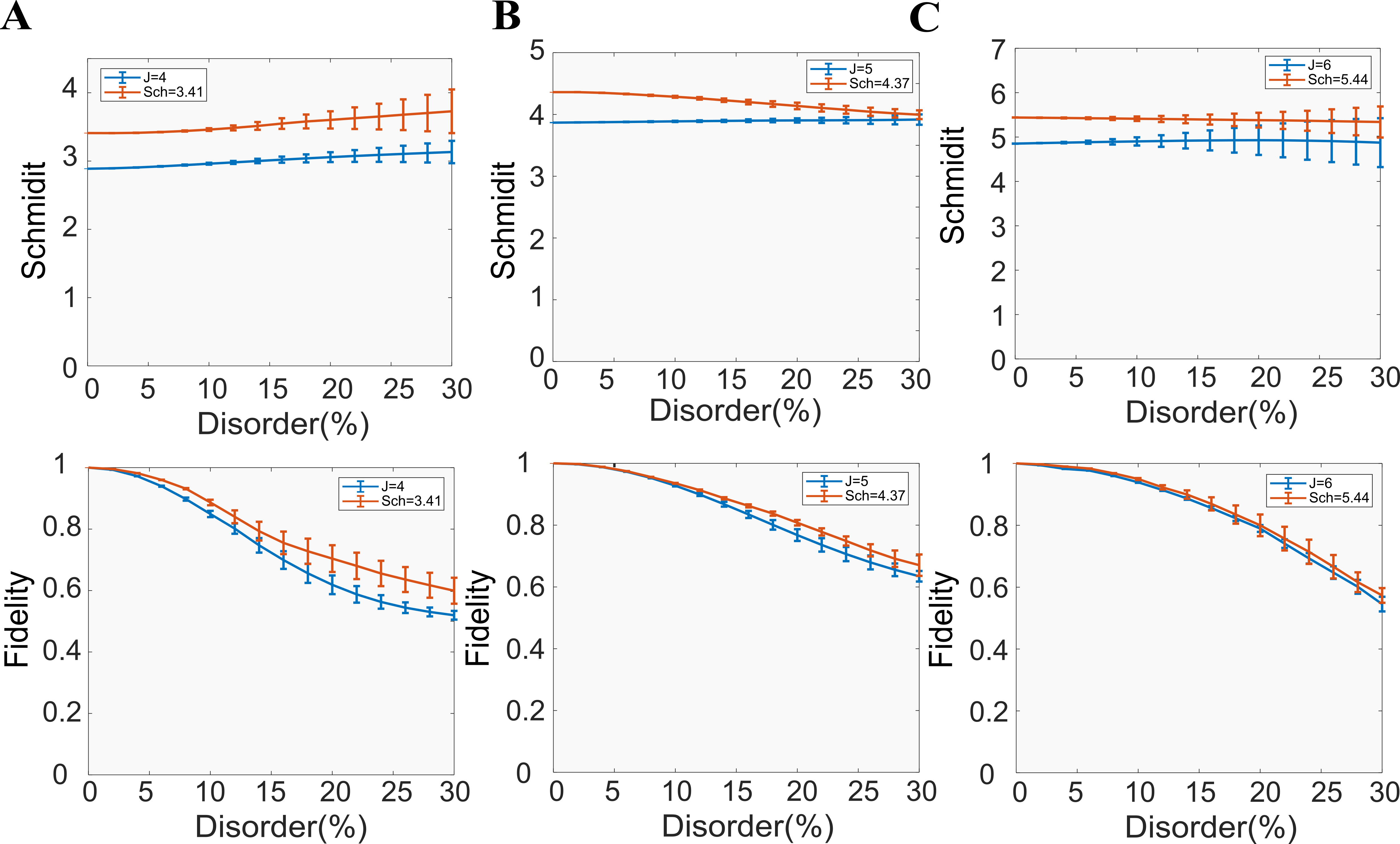}

\caption{\textbf{Disorder resilience of inverse-designed Schmidt-dimensional biphoton states.}
Gaussian-distributed coupling disorder is introduced into the \(J=4\), \(J=5\) and \(J=6\) topological superlattices. 
\textbf{a--c,} Schmidt number, \(K\), as a function of disorder strength for the \(J=4\), \(J=5\) and \(J=6\) lattices, respectively. 
\textbf{d--f,} Corresponding biphoton-state fidelity for the same disorder realizations. 
In each lattice, conventional excitation of the central interface waveguide is compared with the optimized coherent pump superposition designed in the disorder-free device to target \(K_{\mathrm{target}}=3.41\), \(4.37\) and \(5.44\), respectively. 
The optimized pump is kept fixed for all disorder realizations. 
Both excitation schemes retain substantial robustness, reflecting the disorder tolerance of the topological interface manifold. 
The optimized excitation more closely preserves the prescribed Schmidt dimensionality and maintains higher state fidelity as disorder increases, showing that coherent pump engineering reinforces topological protection and enables programmable high-dimensional entanglement under realistic coupling disorder.}

\label{fig:disorder_target_schmidt}
\end{figure}

Figure~\ref{fig:disorder_target_schmidt} shows that the topological superlattice itself provides substantial tolerance to coupling disorder: both the conventional central-waveguide excitation and the optimized pump excitation retain a stable Schmidt number and high biphoton fidelity over a broad disorder range. This confirms that the generated high-dimensional correlations are not fragile features of a perfectly ordered lattice. Importantly, however, the optimized pump input consistently provides an additional level of stability. For \(J=4\), \(J=5\), and \(J=6\), the optimized excitation keeps the Schmidt number closer to the prescribed values \(K_{\mathrm{target}}=3.41\), \(4.37\), and \(5.44\), respectively, while also preserving a higher fidelity under increasing disorder. Thus, pump optimization does not merely tune the entanglement dimensionality in the ideal structure; it selects a coherent excitation of the topological modal manifold that is intrinsically more tolerant to random coupling perturbations.

These results reveal a cooperative mechanism between passive topological protection and active pump-state engineering. The superlattice geometry provides the disorder-resilient modal platform, while the optimized input further stabilizes the desired quantum-state dimensionality and modal overlap. This combination establishes targeted topological-mode excitation as a robust route for programmable generation of high-dimensional biphoton entanglement in realistic photonic lattices.

\clearpage

\bibliography{arXiv}

\end{document}